\documentclass[aps,
prb,
reprint,
showpacs,
amsmath,
longbibliography,
amssymb]{revtex4-1}
\usepackage{graphicx}
\usepackage{dcolumn}
\usepackage{bm}
\usepackage{textcomp}
\usepackage{revsymb}

\begin{document}

\preprint{APS/123-QED}

\title{On atomic structure of  Ge huts growing on the Ge/Si(001) wetting layer
}

\author{Larisa V. Arapkina}
\email{arapkina@kapella.gpi.ru}

\author{Vladimir A. Yuryev}
\altaffiliation[Also at ]{Technopark of GPI RAS} 
\homepage{http://www.gpi.ru/eng/staff\_s.php?eng=1\&id=125}
\email{vyuryev@kapella.gpi.ru}

\affiliation{A.\,M.\,Prokhorov General Physics Institute of the Russian Academy of Sciences, 38 Vavilov Street, Moscow, 119991, Russia}

\date{\today}%

\begin{abstract}

Structural models of growing Ge hut clusters---pyramids and wedges---are proposed on the basis of data of recent STM investigations of nucleation and growth of Ge huts on the Si(001) surface  in the process of molecular beam epitaxy.  It is shown that  extension of a hut base along $<$110$>$ directions  goes  non-uniformly during the cluster growth regardless of its shape. Growing pyramids, starting from the second monolayer, pass through cyclic formation of slightly asymmetrical and symmetrical clusters, with symmetrical ones appearing after addition  every  fourth monolayer. We suppose that pyramids of symmetrical configurations composed by 2, 6, 10, etc. monolayers over the wetting layer are more stable than asymmetrical ones. This might explain less stability of pyramids in comparison with wedges in dense arrays forming at low temperatures of Ge deposition.
Possible nucleation processes of pyramids and wedges on wetting layer patches from identical embryos composed by 8 dimers through formation of 1 monolayer high 16-dimer nuclei different only in their symmetry is discussed. 
Schematics of these processes are presented.
It is concluded from precise STM measurements that top layers of WL patches are relaxed when huts nucleate on them.

\end{abstract}

\maketitle

\section{Introduction}

Ge ``hut'' clusters or small self-assembled Ge/Si(001) clusters faceted by the $\{105\}$ planes and coherent with the substrate lattice (pyramids with square bases  and wedges with rectangular bases elongated in one of the $<$100$>$ directions),\cite{Mo,Iwawaki_SSL,LeGoues_Copel_Tromp,*Eaglesham_Cerullo,*Iwawaki_initial, classification} which form on the Ge wetting layer (WL) at low  temperatures of Si substrates ($\alt 600$\textcelsius) in the process of ultrahigh-vacuum molecular-beam epitaxy (UHV MBE) or---probably with some peculiarities due to hydrogenation of WL---in the process of chemical vapor deposition, have attracted an interest of researchers for more than twenty years since their discovery by Mo \textit{et al.} in 1990\cite{Mo},  because of both their potential  practical importance for development of Si-based monolithic optoelectronic devices\cite{[{See, e.\,g.,} ] Wang-Cha,*Dvur-IR-20mcm,*Dvur-IR-3-5mcm, VCIAN-2012} and convenience and simplicity of their usage as model objects for investigation of the Stranski-Krastanow growth of heteroepitaxial structures.  

A lot of articles published during these years were devoted to a complicated issue of hut appearance and its further growth on the WL (see, e.\,g., Refs.~\onlinecite{Nucleation, *Goldfarb_2005, *Goldfarb_2006, *Goldfarb_JVST-A, Island_growth, Vailionis, Kastner, hut_stability, LLL, Tersoff_LeGoues, Schukin-Bimberg, classification, Hut_nucleation, CMOS-compatible-EMRS} and numerous articles cited therein). 
However, only a few of these works studied this issue on atomic level. \cite{Hut_nucleation, CMOS-compatible-EMRS, VCIAN2011, initial_phase, Nucleation_high-temperatures}  Investigations carried out on atomic-level were mainly devoted to the structure of the \{105\} facets, for which an initially proposed simple model based on paired dimers\cite{Mo} (the so called PD model) was eventually replaced by a model considering step rebonding as a source of the facet stability\cite{Fujikawa,*Fujikawa_ASS} (the latter one is usually referred to as the RS model), rather than to nucleation of the clusters  or their  in-height or longitudinal growth.

\begin{figure*}
\includegraphics[scale=0.15]{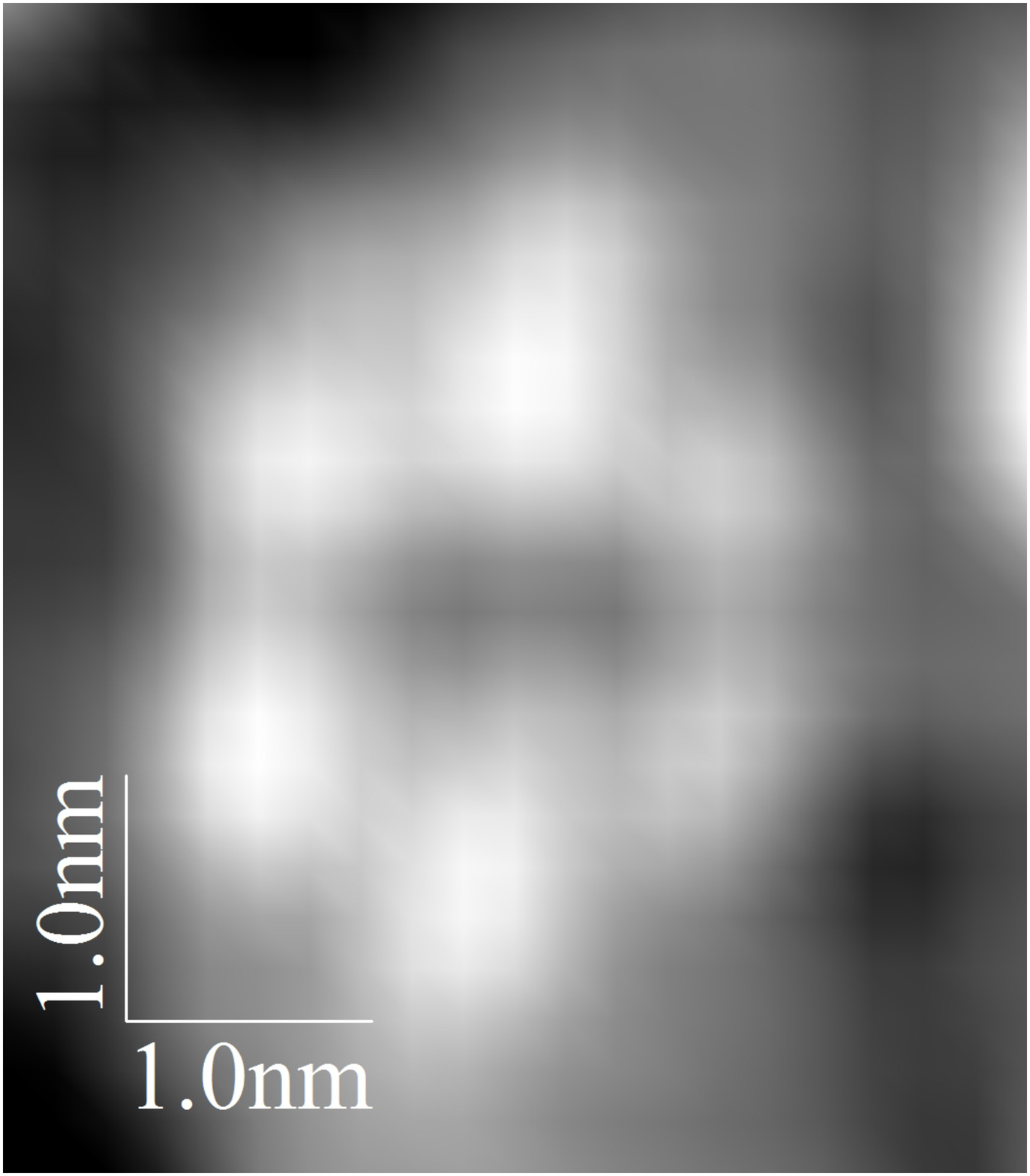}(a)
\includegraphics[scale=0.15]{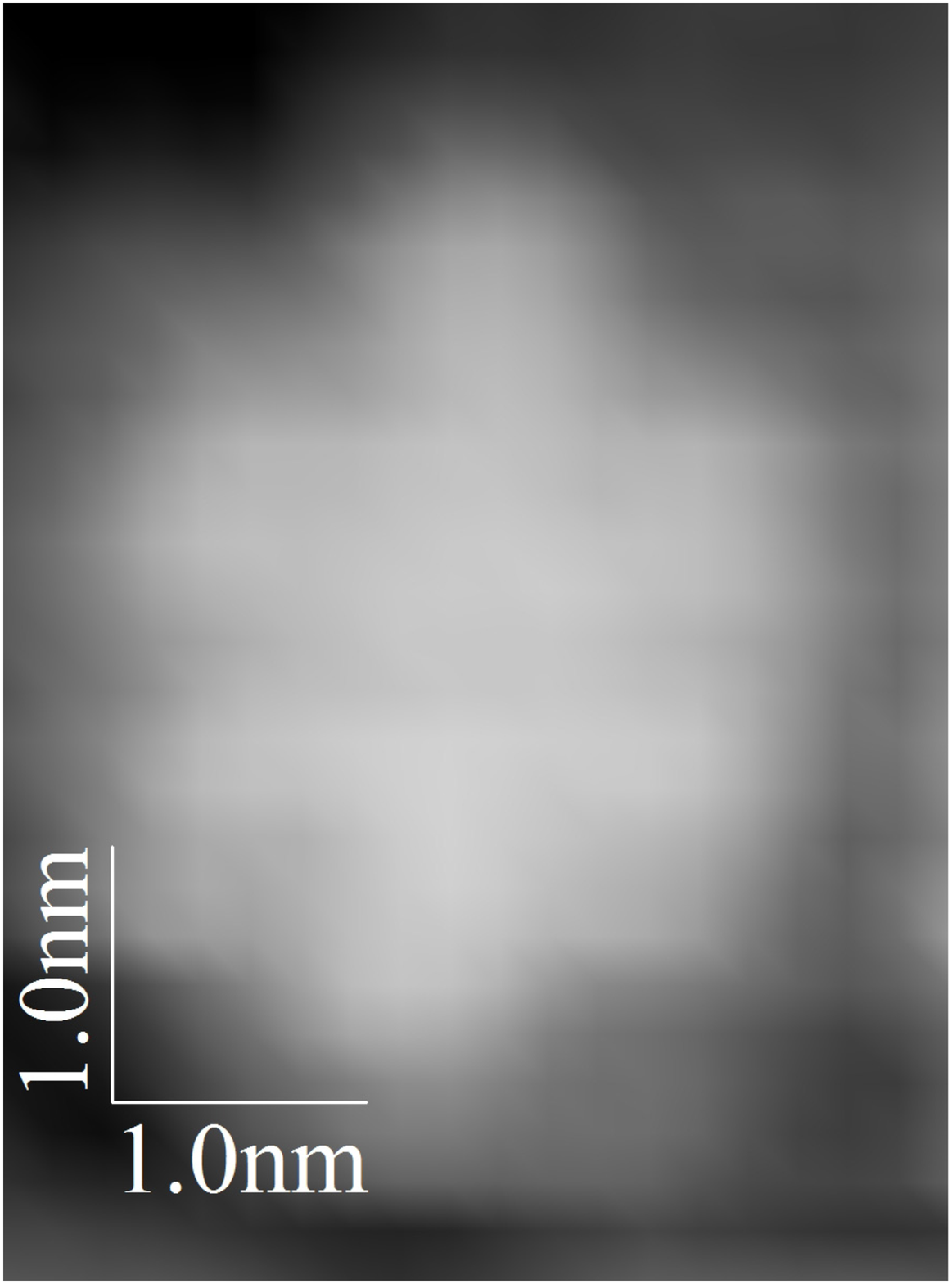}(b)
\includegraphics[scale=.8]{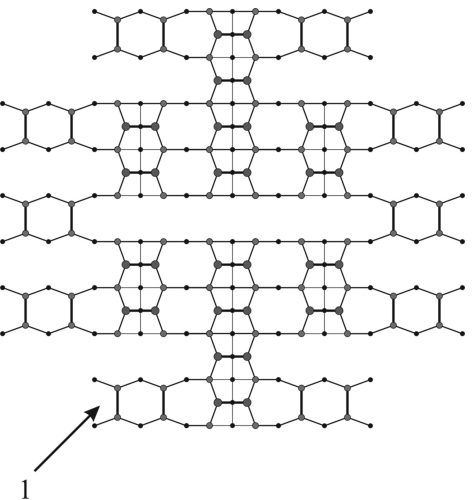}(c)\\
\includegraphics[scale=0.15]{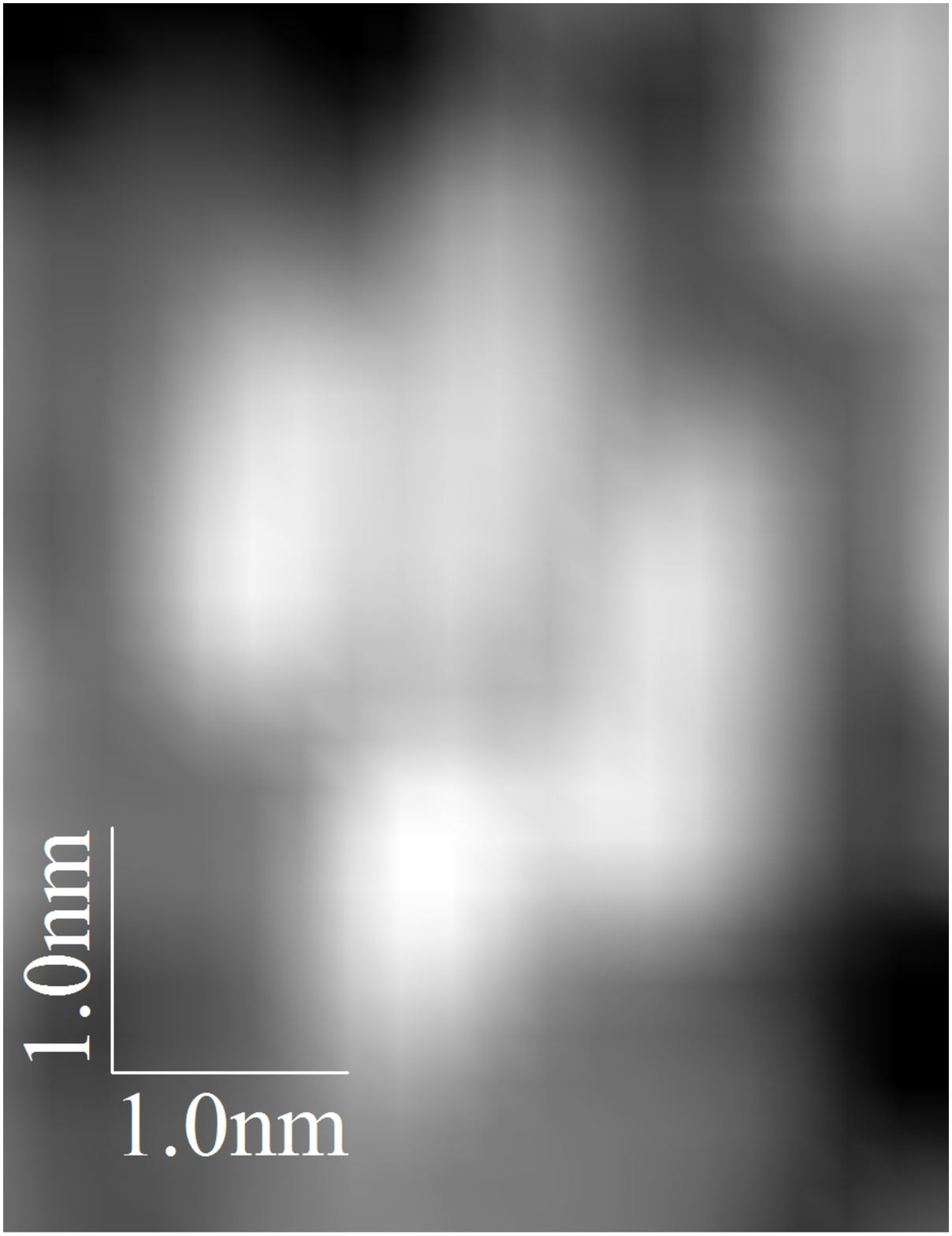}(d)
\includegraphics[scale=0.15]{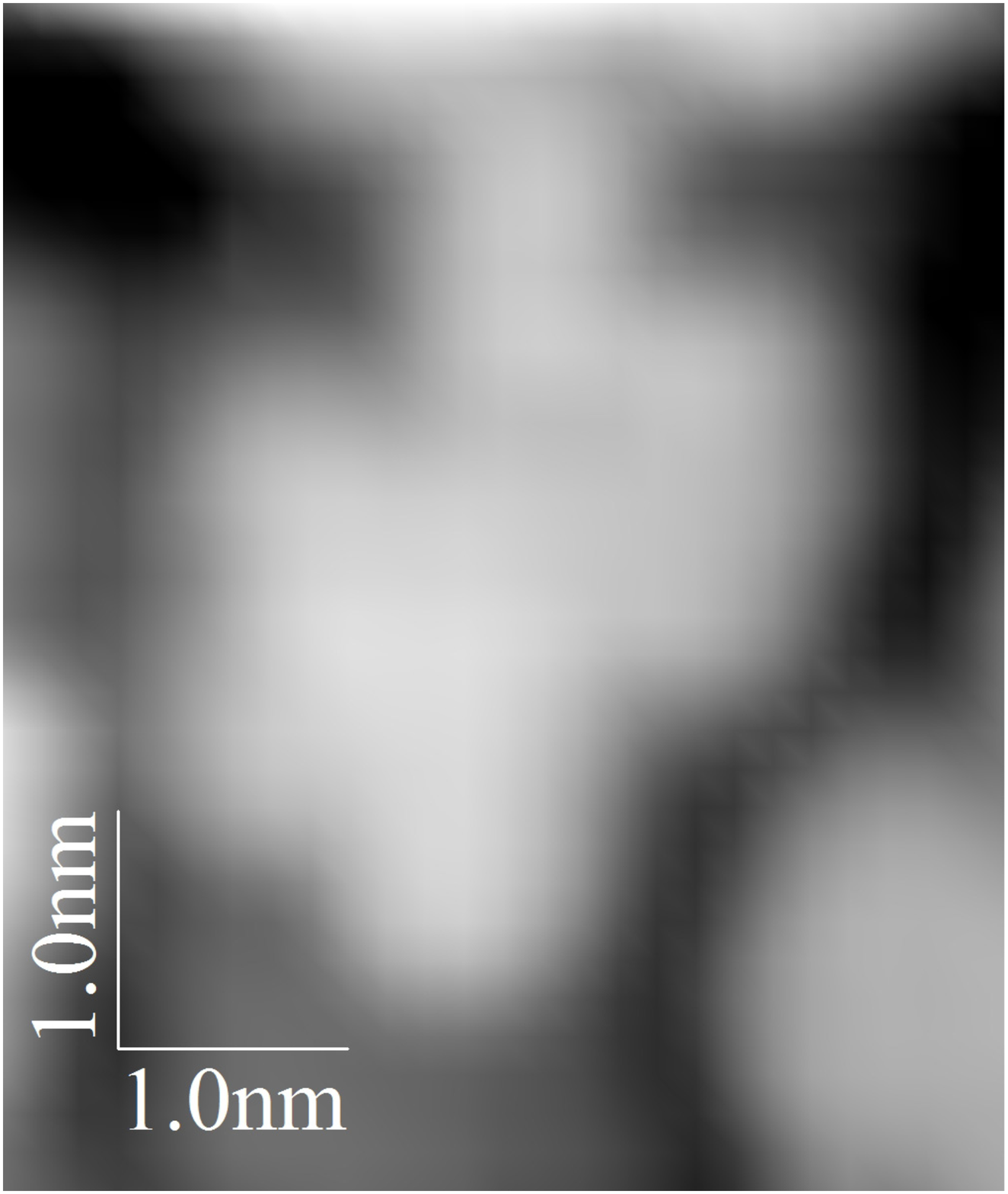}(e)~~~
\includegraphics[scale=0.87]{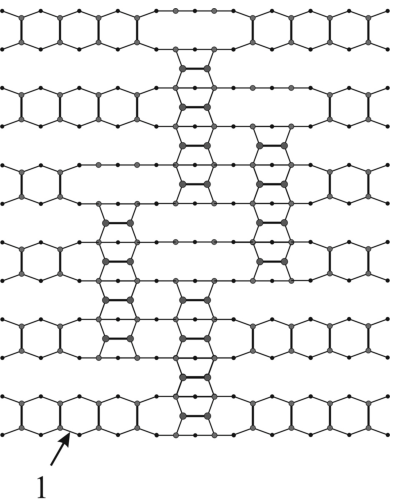}(f)
\caption{\label{fig:nuclei}STM empty state images of Ge hut nuclei on patches of Ge/Si(001) wetting layer and their schematic representation as 16-dimer structures of different symmetry: pyramid nuclei (a,\,b) and a sketch of their atomic structure (c); wedge nuclei (d,\,e) and their structure (f);\cite{Hut_nucleation} figure `1' designates dimer rows of the top layer of a wetting layer patch.
}
\end{figure*}

Our recent experimental explorations\cite{Hut_nucleation,CMOS-compatible-EMRS, VCIAN2011, initial_phase, Nucleation_high-temperatures, atomic_structure, VCIAN-2012} carried out on atomic level by high-resolution STM have demonstrated  fine details of hut nucleation and its evolution during the growth which however have not been sufficiently interpreted thus far and presented in terms of structural schematics and drawings which would be helpful for further theoretical calculations and numerical simulations. For example, we have discovered the phenomenon of simultaneous appearance of two types of nuclei\cite{Hut_nucleation,Nucleation_high-temperatures} which are composed by 16 dimers and different only in symmetry---a separate nucleus for each species of huts---on tops of Ge WL $M\times N$ patches\cite{[{See, e.\,g.,} ] Chem_Rev,*Wetting} of 4-monolayer\cite{initial_phase} (ML) height (Fig.~\ref{fig:nuclei}). Unfortunately, no satisfactory explanation of this phenomenon has been proposed thus far which would describe a driving force of this strange behaviour of Ge dimers which group in two different formations,\cite{initial_phase,Nucleation_high-temperatures} instead of a single one, 
on tops of 
WL patches after the latter have reached their maximum thickness.\footnote{
We should notice here that Ge/Si(001) WL patches are well isolated from one another by a grid of dimer vacancy trenches\cite{Chem_Rev,Wetting} (``vacancy rows'' and ``row vacancies''): this statement is grounded on our observations of simultaneous but independent formation of the $c(4\times 2)$ and  $p(2\times 2)$ reconstructions on tops of adjacent WL patches as well as on observation of appearance of a pyramid nucleus and a wedge nucleus on neighbouring WL patches.\cite{initial_phase} So, a process of hut nucleation should not be considered as formation of a 3D relief on a more or less homogeneous elastically strained infinite thin film; for adequate description, it should be treated as reconstruction of separate (isolated or weakly interacting) WL patches via formation of specific atomic structures on their tops. Effect of patch boundaries and area (dimensions) on this process, as well as its top layer reconstruction, should be taken into account. (See also Sec.\,\ref{sec:WL_structure}.)

A detailed thermodynamic analysis of coherent islands formation on mismatched WL can be found in the review article by Schukin and Bimberg.\cite{Schukin-Bimberg} Microscopic consideration of atomic-level processes resulting in appearance  of hut nuclei on WL patches and their further expansion to 3D huts is out of the scope of that paper, however} 

We have demonstrated  that both pyramids and wedges grow in height conserving the width and the atomic structure of their apexes (the topmost (001) terraces).\cite{classification, VCIAN2011,VCIAN-2012} This phenomenon also requires comprehension and adequate description: up to date, we have only hinted at such description in our recent articles.\cite{Hut_nucleation,CMOS-compatible-EMRS,VCIAN2011}

An aim of this article is to propose physical processes on WL and cluster facets, and their structural schemes, on the basis of the previously performed STM studies, which could give adequate interpretation on atomic level to the experimentally observed evolution of each species of hut clusters during their growth setting the direction of thought for explanation of the presently incomprehensible phenomenon of simultaneous appearance of two types of hut nuclei on Ge/Si(001) WL which give rise to two mutually independent species of huts.\cite{classification,Hut_nucleation}

\section{Experimental}

\subsection{Techniques}

Experiments were carried out using a specially built setup\cite{CMOS-compatible-EMRS,VCIAN2011} consisting of a UHV MBE vessel (Riber EVA~32) connected with a UHV STM chamber (GPI~300)\cite{STM_GPI-Proc,*gpi300}. Details of the pre-growth treatments of Si wafers, which included chemical etching and oxide removal by short high-temperature annealing ($T\sim$~900\textcelsius), can be found in our previous articles cited in Refs.~\onlinecite{our_Si(001)_en, stm-rheed-EMRS}. Thin films of Ge were deposited directly on the clean Si(001) surface, purified from the oxide,\cite{stm-rheed-EMRS} at the temperatures of 360, 530, 600 or 650\textcelsius. Parameters of Ge deposition processes as well as results of our structural explorations of Ge huts and wetting layer performed by high-resolution UHV STM are presented in detail, e.g., in Refs.~\onlinecite{Hut_nucleation,atomic_structure, initial_phase,CMOS-compatible-EMRS,classification,Nucleation_high-temperatures,VCIAN2011,VCIAN-2012}. The WSxM software was used for processing of STM images.\cite{WSxM}

\subsection{Summary of Main Results}

Beginning the presentation of ideas and conclusions, which are in the focus of the current article, we would like to outline for the readers the main experimental observations and models on which our further consideration is based.

\begin{figure*}
\includegraphics[scale=.8]{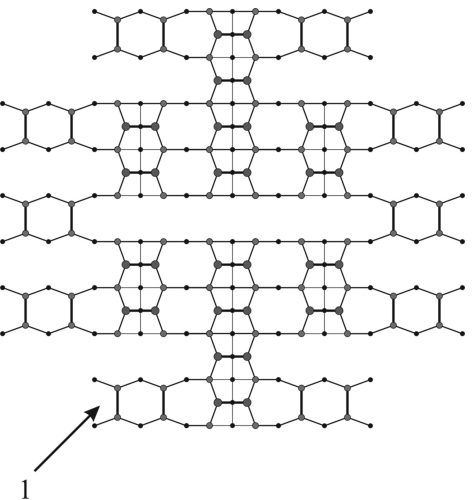}(a)
\includegraphics[scale=.9]{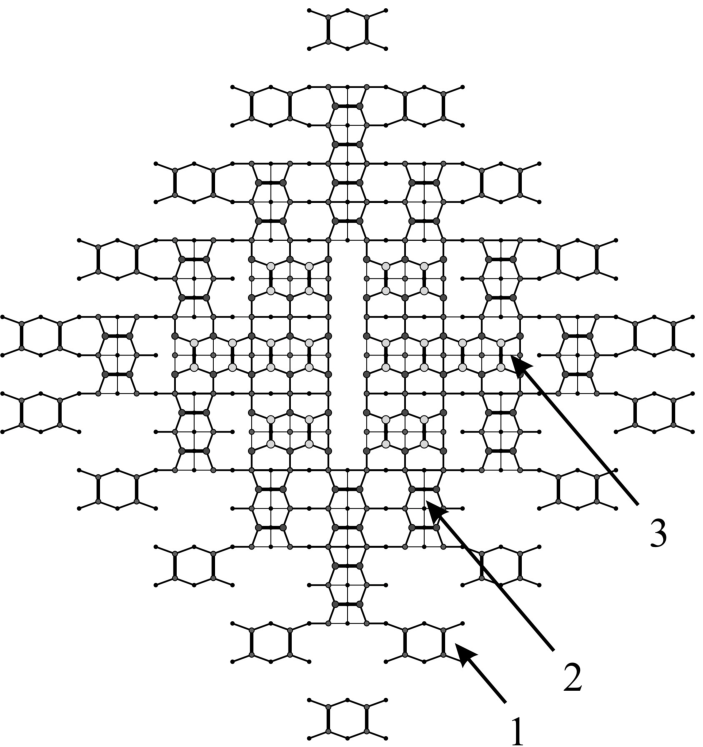}(b)\\
~\\
\includegraphics[scale=.82]{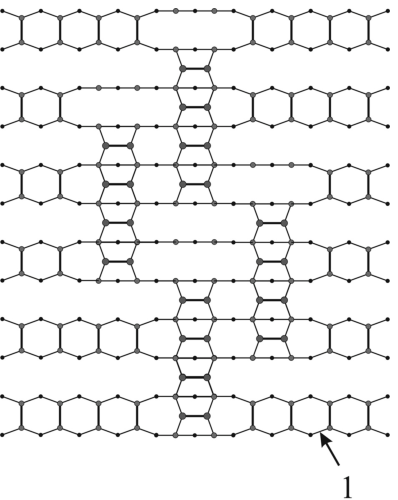}(c)
\includegraphics[scale=.8]{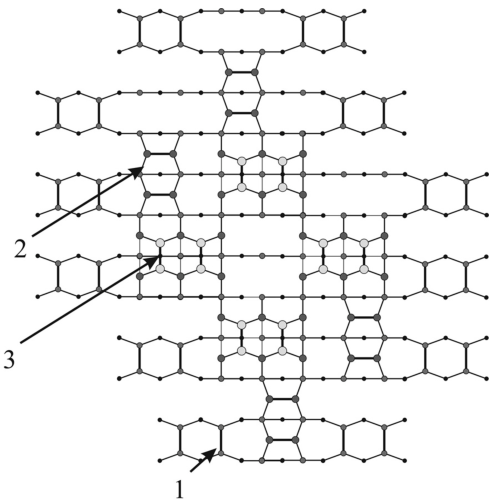}(d)
\includegraphics[scale=1.05]{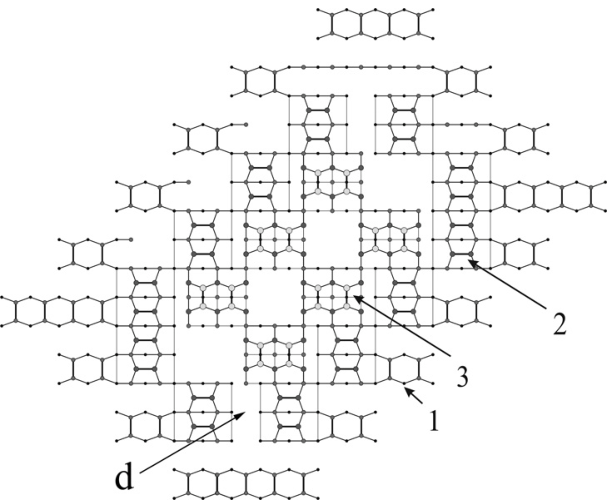}(e)
\caption{\label{fig:reconstruction}Schematics of hut nucleation: a nucleus of a pyramid  arising reconstructs the surface a WL patch (a) and then transforms into a 2-ML pyramid (b) which repeats the nucleus structure on its apex; a nucleus of a wedge also reconstructs the WL patch surface (c) and the second layer reconstructs the nucleus to form a correct structure of its apex (d); a 2-ML wedge repeats this structure on its ridge (e); when a wedge  takes the right shape, point defects  arise on its opposite triangular \{105\} facets (the defect is shown by the letter `d' on one of the triangular facets) and determine the direction along the particular  $<$100$>$ axis for cluster elongation;\cite{Hut_nucleation} figures `1', `2' and `3' indicate WL, first and second terrace of the clusters, respectively.
}
\end{figure*}

It has been  known also since the pioneering works\cite{Mo,Iwawaki_SSL} that facets of huts are (105) planes formed by (001) terraces  separated by monoatomic steps and that clusters grow in height due to formation of new (001) terraces.  But their exact atomic structure has been disputable.\cite{Fujikawa,*Fujikawa_ASS} Our recent works\cite{Hut_nucleation,atomic_structure, CMOS-compatible-EMRS} have presented additional data of an accurate STM study of the hut facet structure. The obtained STM micrographs of the \{105\} facets of huts have been shown to correspond to the PD model\cite{Mo} rather than the RS\cite{Fujikawa,Fujikawa_ASS} one, and a conclusion has been made that \{105\} facets of huts consist of non-rebonded (001) terraces;\cite{atomic_structure} the same conclusion has been made from a simple crystallographic consideration of nucleating huts (Figs.~\ref{fig:nuclei} and \ref{fig:reconstruction}).\cite{Hut_nucleation,CMOS-compatible-EMRS}
The width of (001) terraces on the \{105\} facets has been found to be equal to 2   translations of the crystalline lattice in the $<$110$>$ direction. 

As mentioned above, at the growth temperatures below 600\textcelsius\, formation of two species of Ge hut clusters---ones with square bases or pyramids and ones  with rectangular bases elongated in one of the $<$100$>$ directions or wedges---is observed on WL.\cite{Hut_nucleation,classification,Mo, Kastner,Pchel_Review-TSF} Our STM data demonstrate these two cluster forms to grow from different types of nuclei which have different structures and symmetries:\cite{Hut_nucleation,initial_phase} Fig.~\ref{fig:nuclei} presents STM images of these nuclei and models of arrangement of their atoms on WL. Exploration of apex structure of bigger huts has shown that it is also different for pyramids and wedges.\cite{classification,Hut_nucleation} Both nuclei reconstruct WL patch tops on which they appear (each on a single patch).  Nuclei of pyramids transform to 3-D huts without any further reconstruction; a growing pyramid always  reproduces a blossom-like shape of the nucleus on its vertex keeping its apex unchanged (Fig.~\ref{fig:reconstruction}\,a,\,b).\cite{Hut_nucleation} Nuclei of wedges are being reconstructed when the second layer of a cluster forms (Fig.~\ref{fig:reconstruction}\,c,\,d);\cite{classification,Hut_nucleation} as a result of this shape transition a formation arises  which then is repeated as a basic unit in the structure of a ridge of a wedge-like cluster of any height and width (Fig.~\ref{fig:reconstruction}\,d,\,e). Thus, the width of apexes of wedges of any height and width is also permanent.\cite{classification,Hut_nucleation}    
Meanwhile, as it follows from the presented clear structural model, point defects form at the penultimate terrace of both  triangular facets of a wedge  because of the ambiguity of locations of the dimer pairs near the gap indicated by the letter `d' in Fig.~\ref{fig:reconstruction}\,e. This pair of defects on the opposite sides of a cluster likely determines the $<$100$>$ axis, along which the cluster can then  elongate, by removing the degeneracy of its just nucleated facets composed by only two narrow terraces and two monoatomic steps. Imagine that the island in Fig.~\ref{fig:reconstruction}\,e is shorter so that its topmost terrace is composed of only two couples of dimer pairs; in this case the island is nearly square-based (it resembles a pyramid) but its symmetry is however violated by the presence of the defects, the degeneracy of facet energies  is removed due to difference in stress of `triangular' and `trapezoidal' facets caused by the defects and the $<$100$>$ direction of elongation is set. 
It seems very likely that namely the  existence of these defects is a reason of the longitudinal growth of wedges\cite{Hut_nucleation,CMOS-compatible-EMRS}  resulting in formation of very long huts\cite{classification} and even quantum wires of enormous\cite{Ge_nanowires_Zhang} aspect ratios.  Notice, that real pyramids (Fig.~\ref{fig:reconstruction}\,a,\,b) do not contain such defects on any pair of their opposite triangular facets and, likely, this is the reason that they cannot elongate: their symmetry is undisturbed and their facets are energetically degenerate even at very beginning of formation.  This is one of the most important structural differences of pyramids and wedges which likely determines the difference in the processes of their growth. This explains why the shape transitions between clusters are prohibited and each species of huts keeps its identity during growth. 

A commonly adopted opinion related to occurrence of wedges (elongated huts) is that they arise due to elongation of pyramids in one of two equivalent $<$100$>$ directions;\cite{Tersoff_Tromp,Goldfarb_2006,Island_growth}  an issue of preferential growth on one (or on two opposite) of four energetically degenerate facets is quite unclear in such an approach, however.\cite{classification}
As we mentioned above, according to the data of our STM studies and to corresponding structural models such transition is impossible, a pyramidal hut cannot change its symmetry and transform into wedge-like one (and vice versa): apexes of these clusters, as well as their nuclei and, as a consequence, triangular facets, have different structures and the in-height growth appears to be the only possible way of expansion for pyramids.\cite{classification, Hut_nucleation, CMOS-compatible-EMRS}

\section{Hut growth and appearance: Models and Processes}

Now, being aware of the necessary empirical facts and their interpretation, we can proceed with consideration of atomic models of processes which explain them. 

\subsection{Models of uniform and non-uniform growth}

\begin{figure*}
\includegraphics[scale=.1]{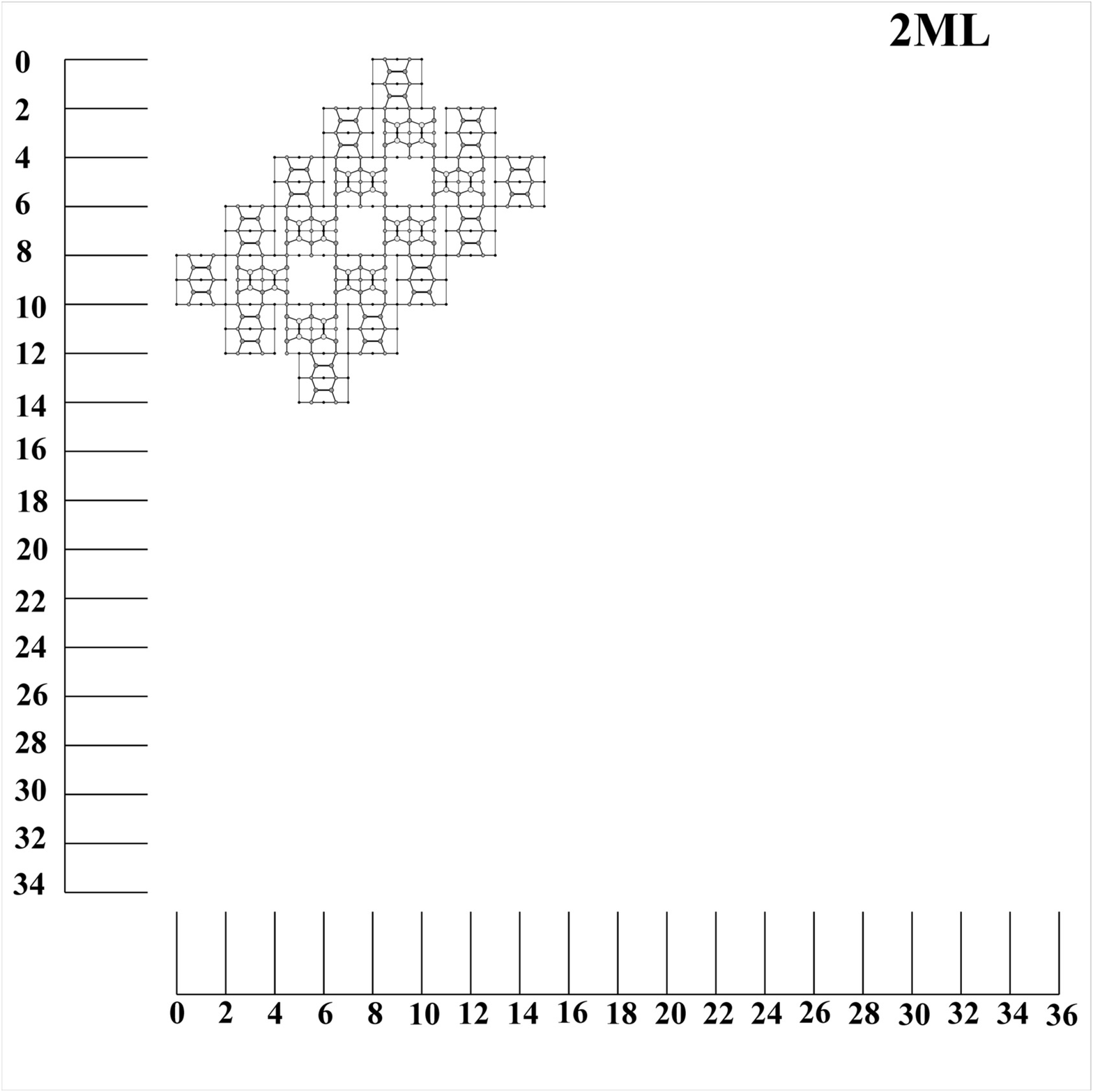}(a)
\includegraphics[scale=.1]{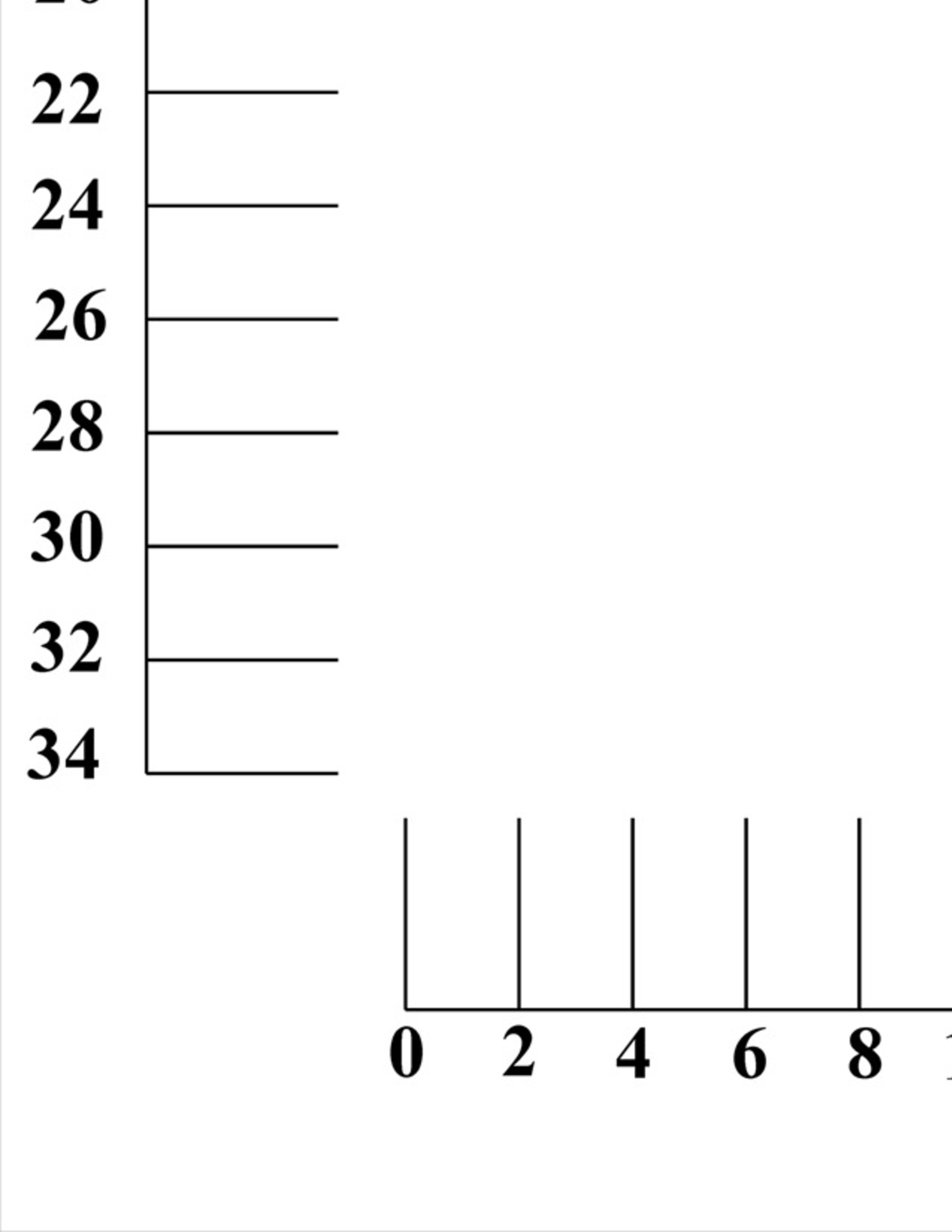}(b)\\
\includegraphics[scale=.1]{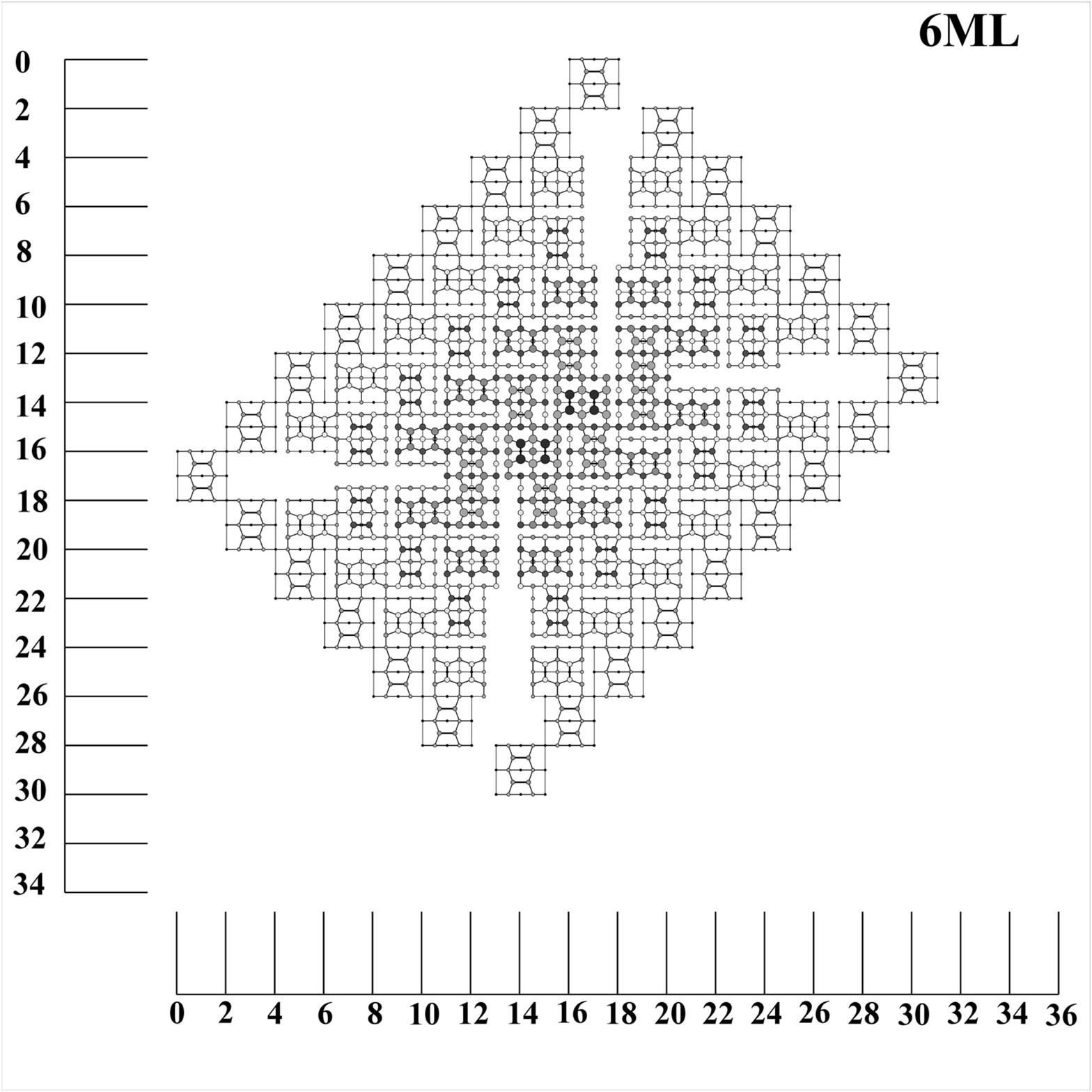}(c)
\includegraphics[scale=.1]{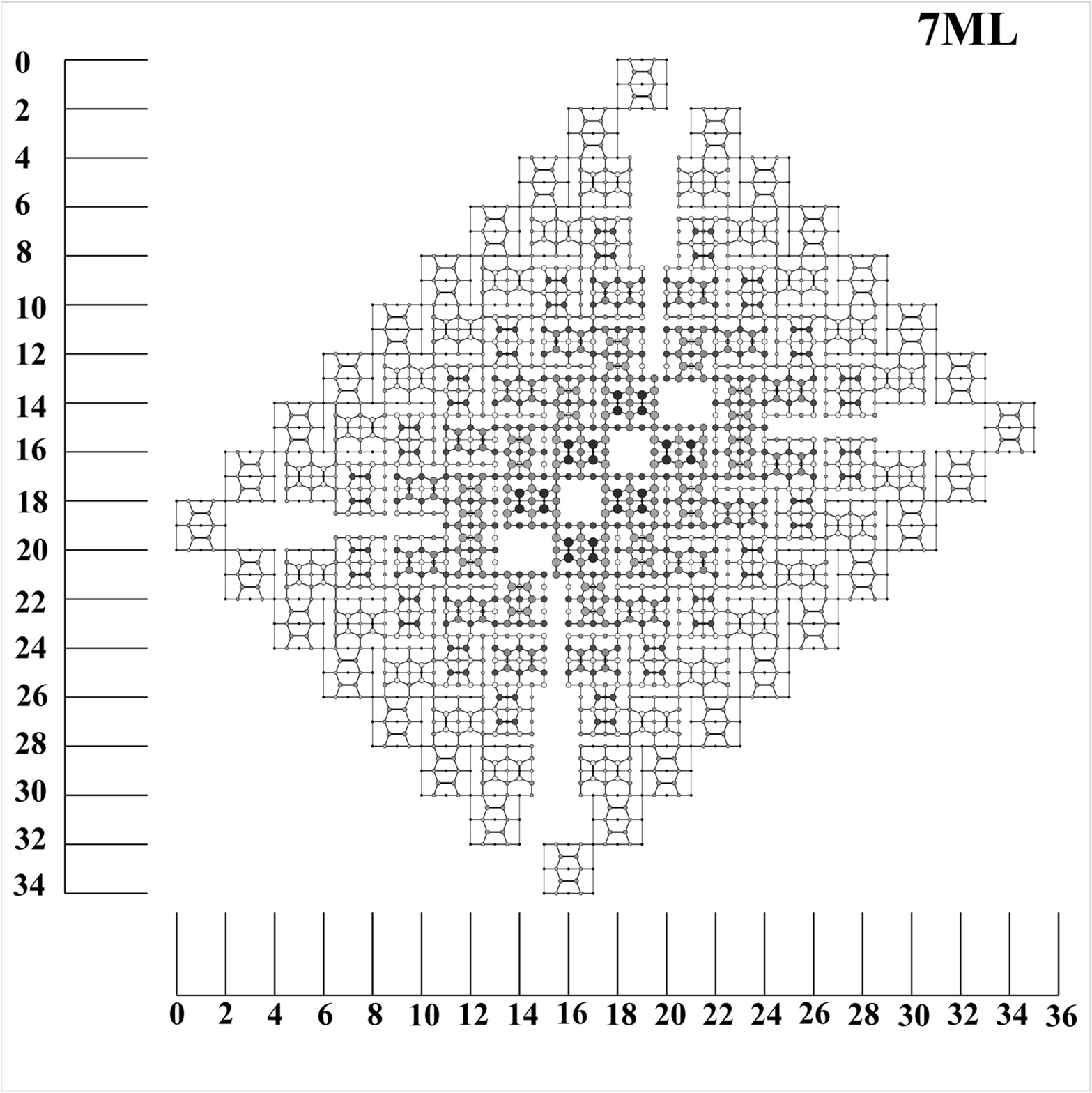}(d)
\caption{\label{fig:uniform_wedge}
A model of uniform growth of a wedge-like hut: the cluster effective height over WL is 2 (a), 3 (b), 6 (c) and 7 (d) ML; the real height of the latter hut is only 6\,ML over WL but its width corresponds to 7\,ML. Only the structures of the \{105\} facets and apexes are shown. The scales show the cluster base widths along the $<$110$>$ axes expressed as a number of elementary translations.
}
\end{figure*}

\begin{figure*}
\includegraphics[scale=.1]{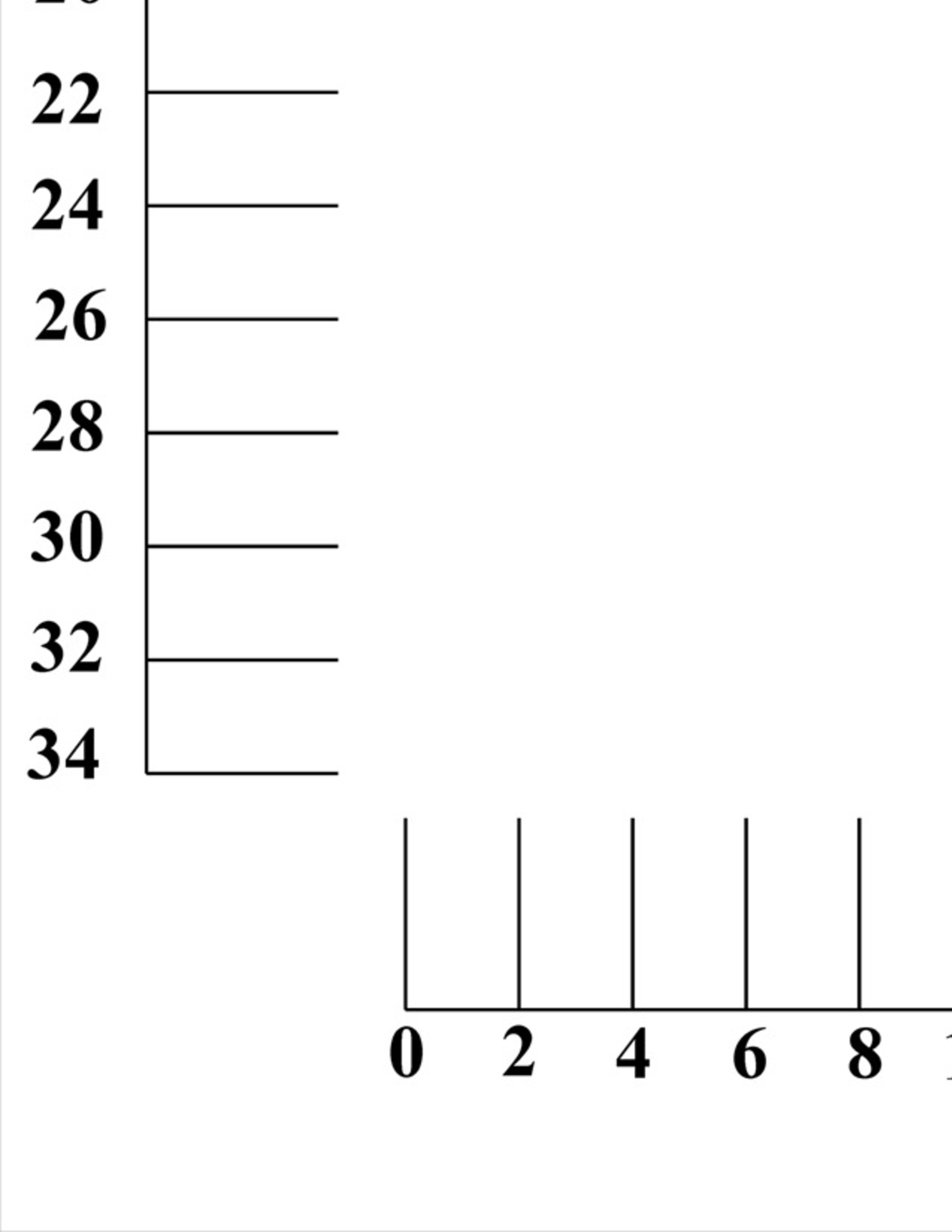}(a)
\includegraphics[scale=.1]{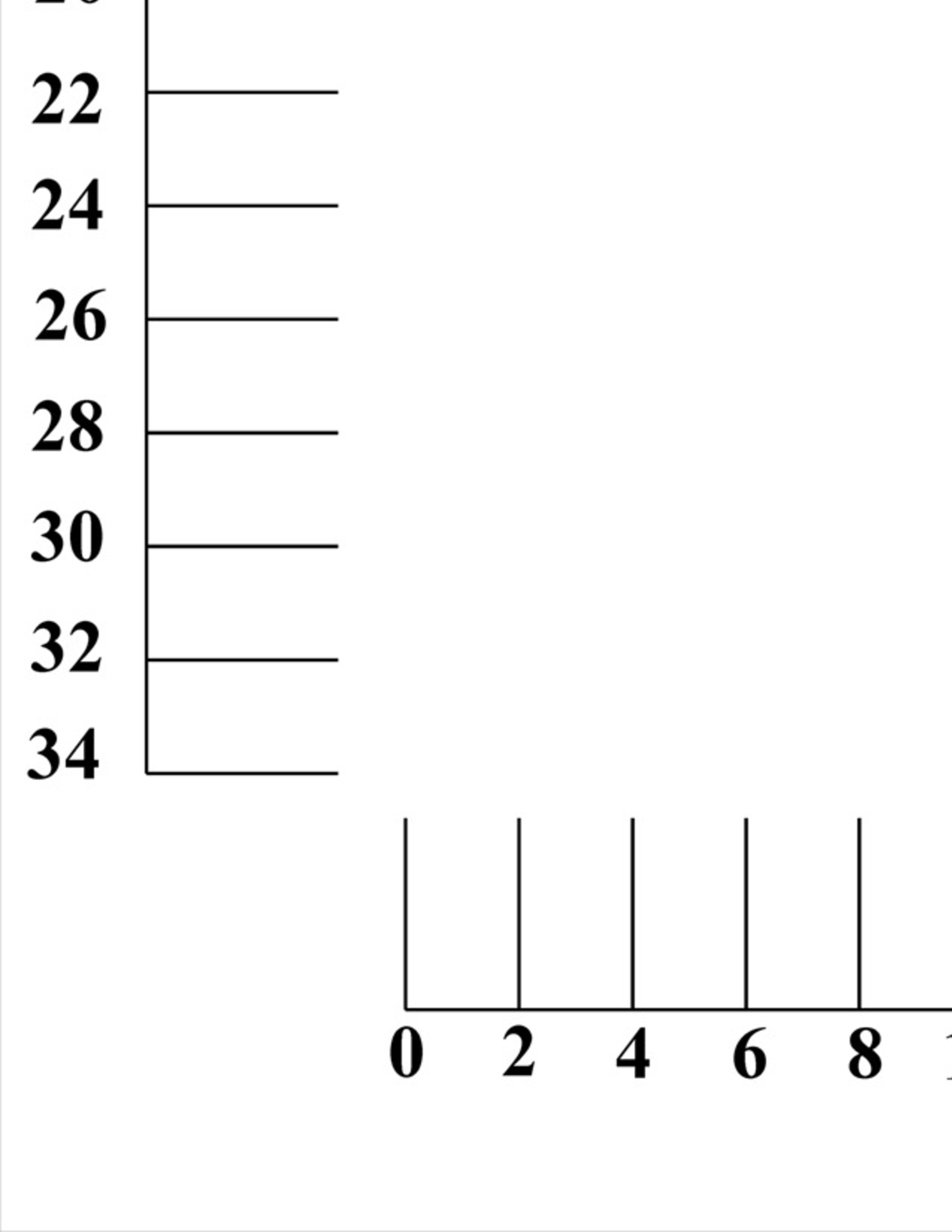}(b)\\
\includegraphics[scale=.1]{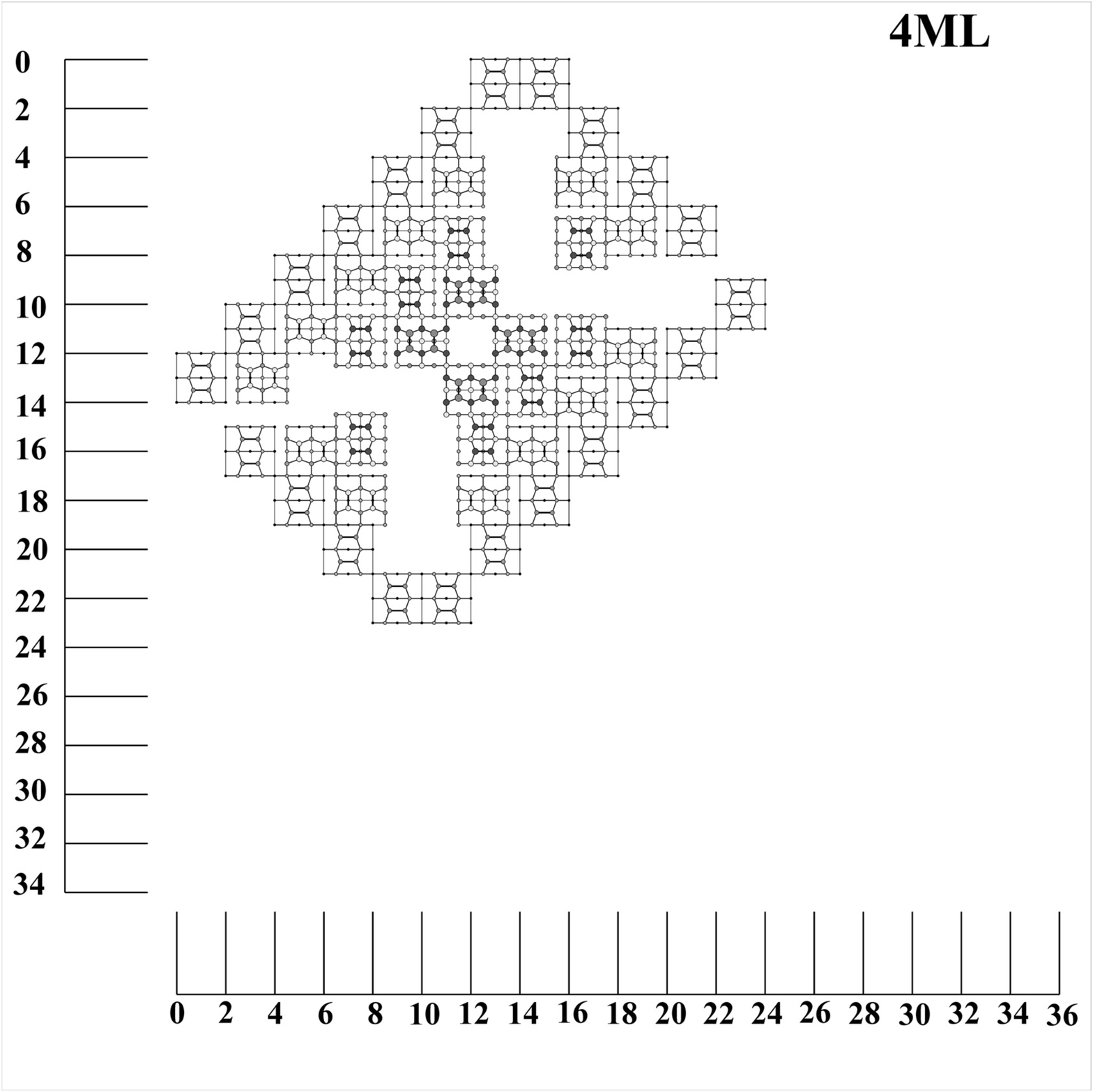}(c)
\includegraphics[scale=.1]{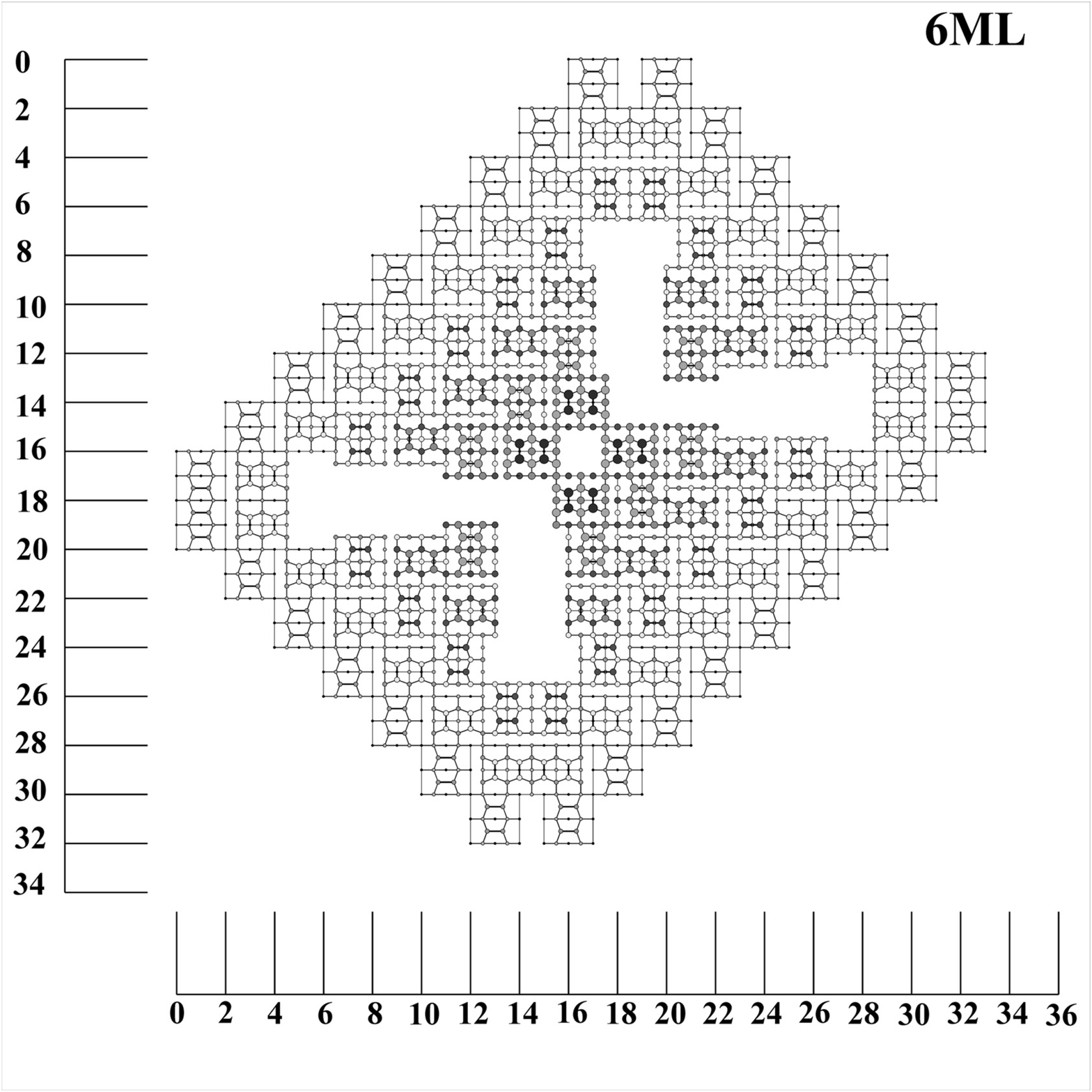}(d)\\
\includegraphics[scale=1.7]{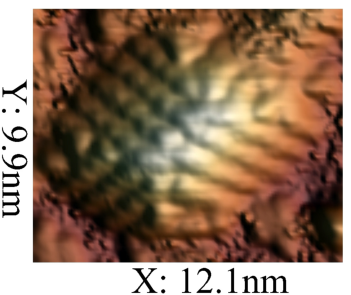}(e)~~~~~~~~~~~~~~~~~~~~~~~~~~~~~~~~~~~~~~~~~~~~~~~~~~~~~
\caption{\label{fig:non-uniform_wedge}(Color online) 
A model of non-uniform growth of a wedge-like hut: the number of completed layers over WL is 2 (a), 3 (b), 4 (c) and 6 (d). Only the structures of the \{105\} facets and a unit cell of the apexes are shown. The abscissa and ordinate axes show the cluster base widths along the $<$110$>$ axes expressed as a number of elementary translations. An STM micrograph of a wedge (4\,ML high over WL) is given as an example demonstrating the apex structure [Ge deposition temperature is 360\textcelsius, Ge coverage is 5.4\,\AA, bias voltage is +1.8\,V, tunneling current is 100\,pA].
}
\end{figure*}

\begin{figure*}
\includegraphics[scale=.5]{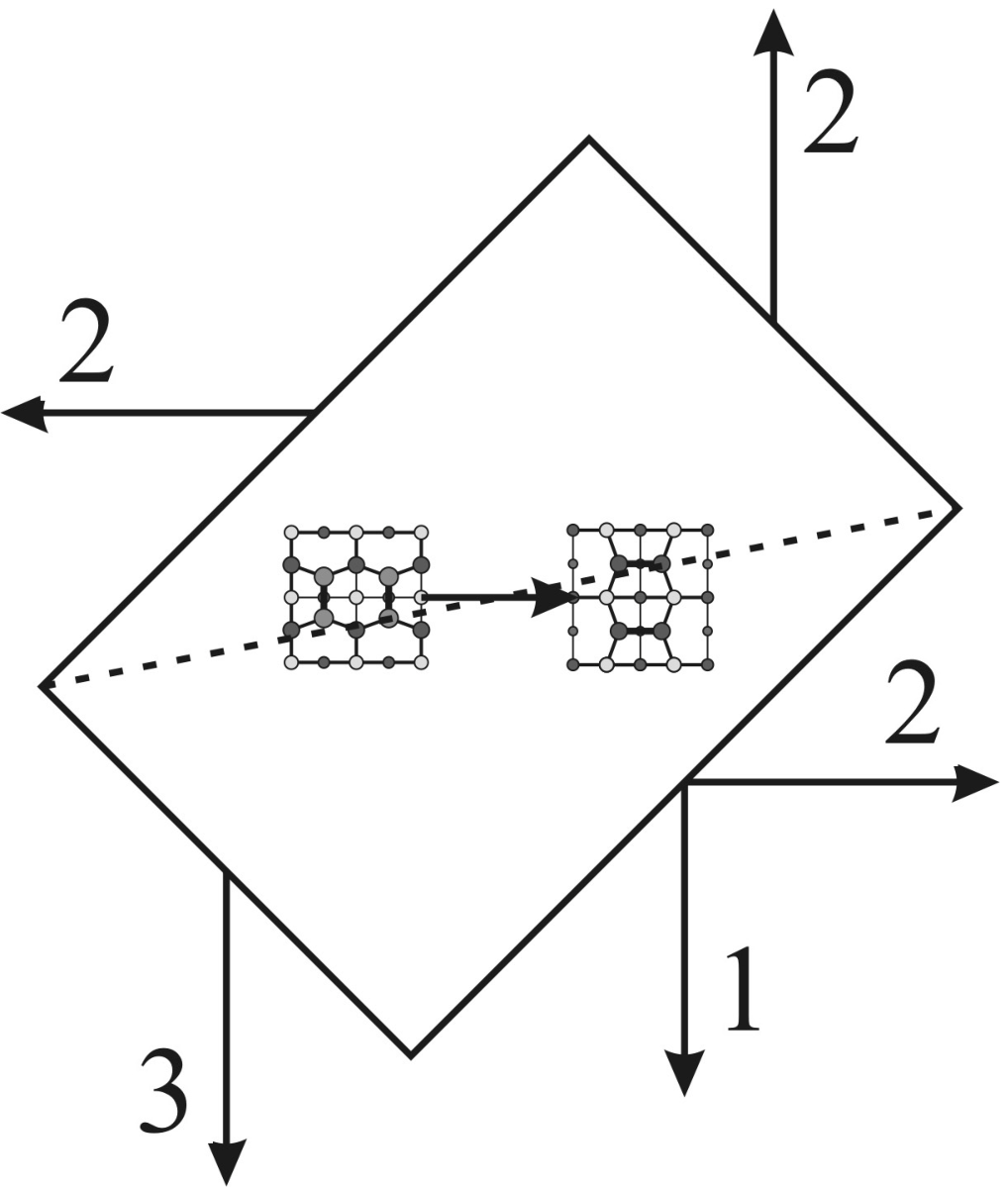}(a)
\includegraphics[scale=.5]{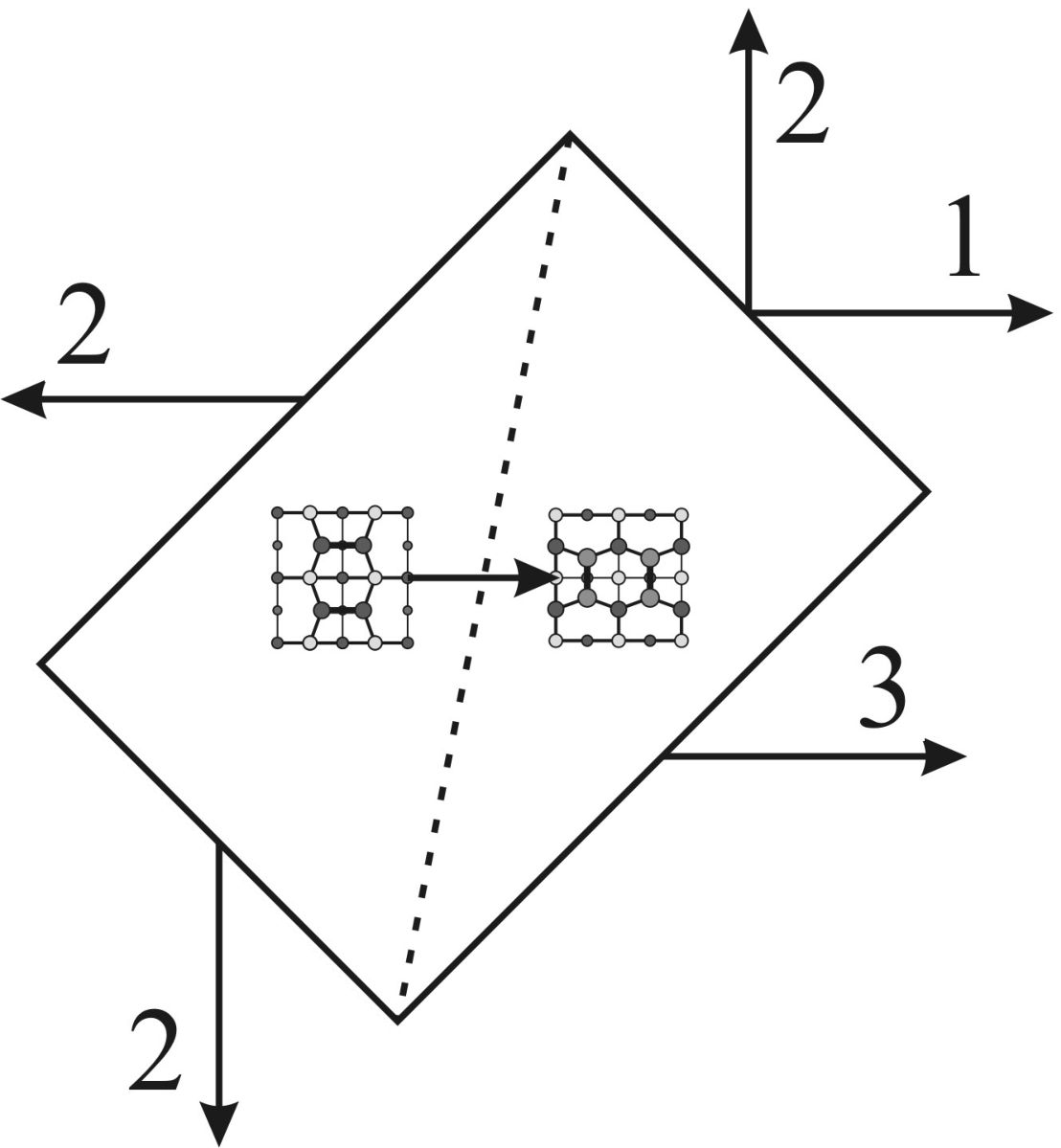}(b)
\caption{\label{fig:wedge}
Schematic representation of non-uniform growth by 1\,ML of a wedge-like hut-cluster depending on the initial and final direction of dimer rows on its apex: the apex transformation is shown by sketches of atomic configurations of the unit cells connected by arrows on the diagonals of the rectangles; the arrowheads show the final directions. Arrows at bases show the $<$110$>$ directions and figures near them designate a number of elementary translations by which the corresponding base side is shifted in each direction due to the increase in the cluster height by 1\,ML. 
}
\end{figure*}

\begin{figure*}
\includegraphics[scale=.1]{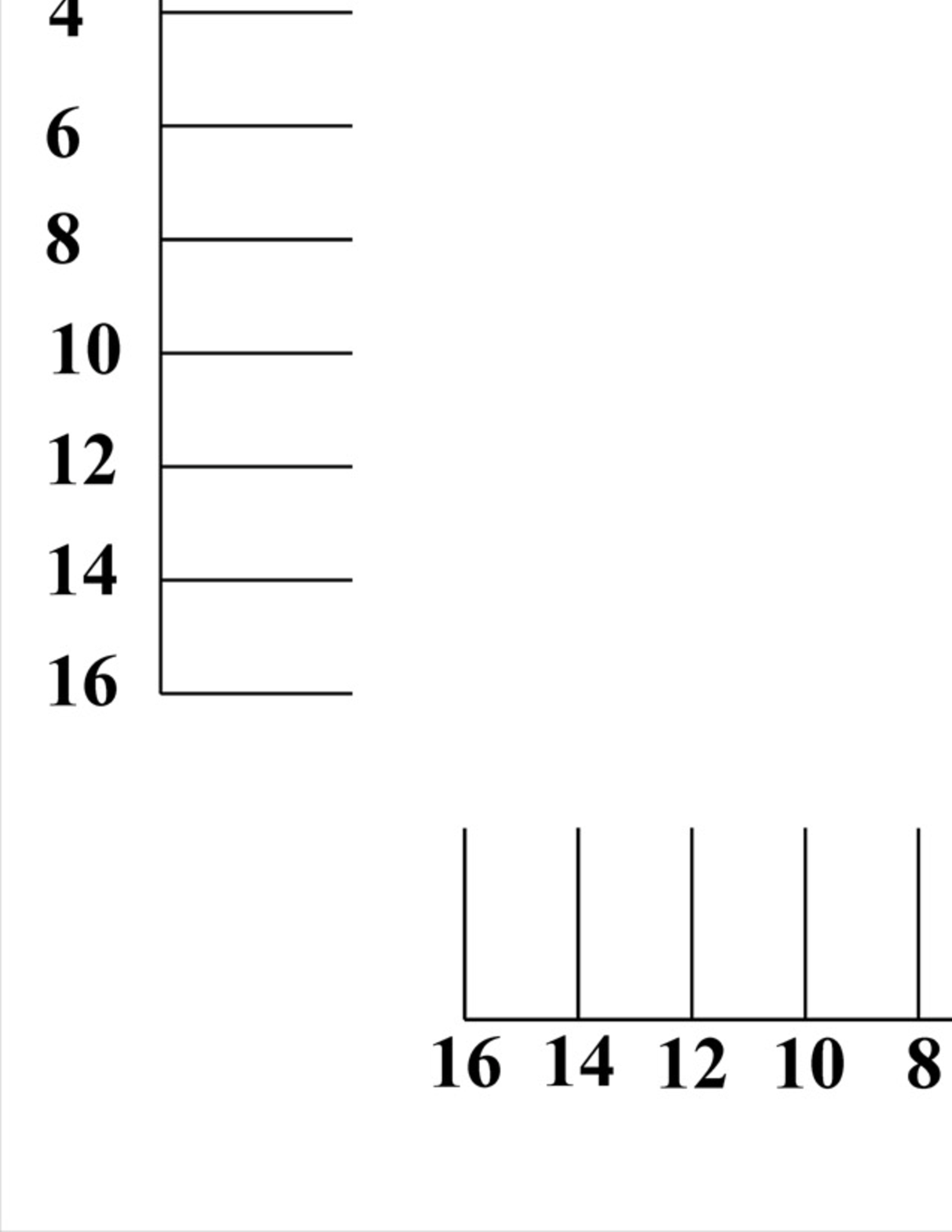}(a)
\includegraphics[scale=.1]{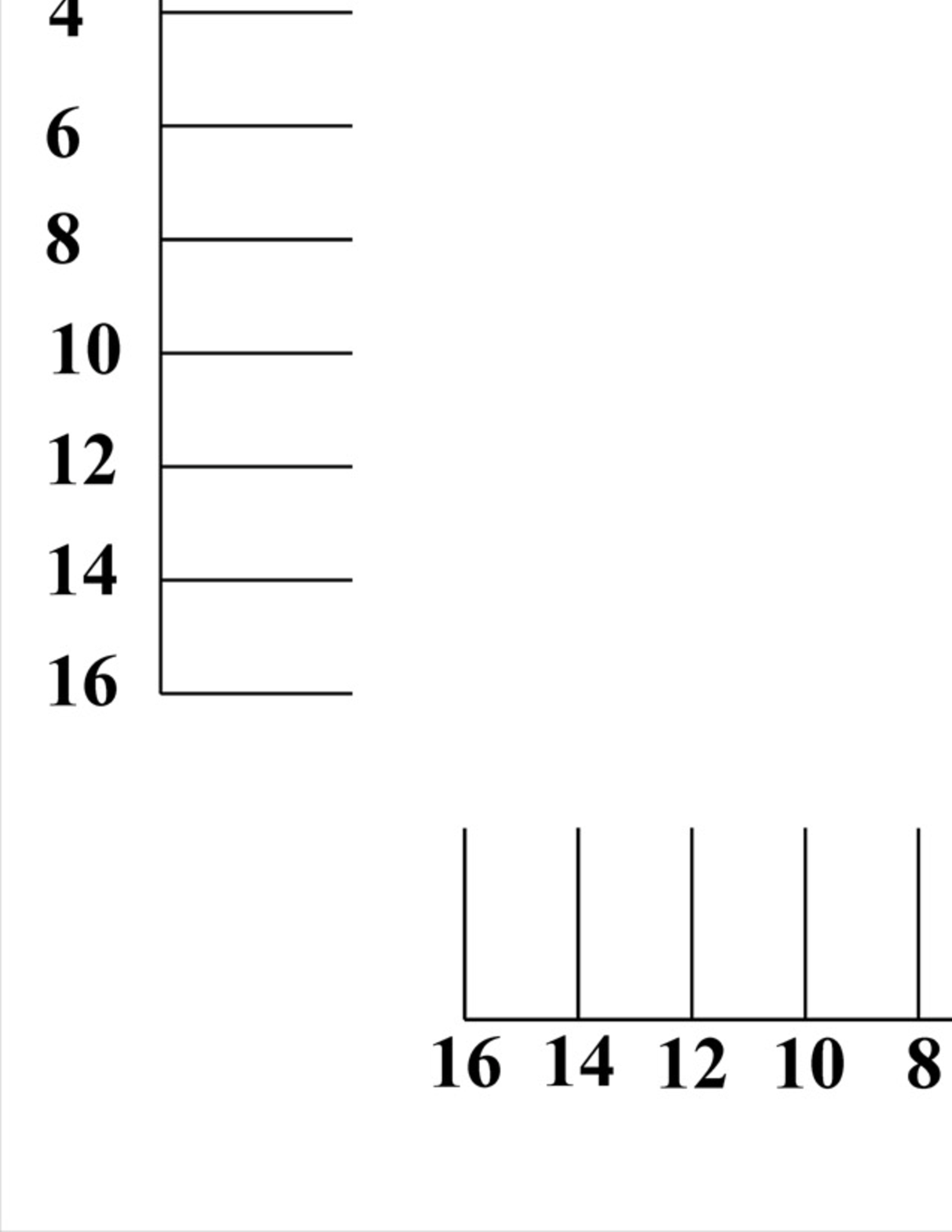}(b)\\
\includegraphics[scale=.1]{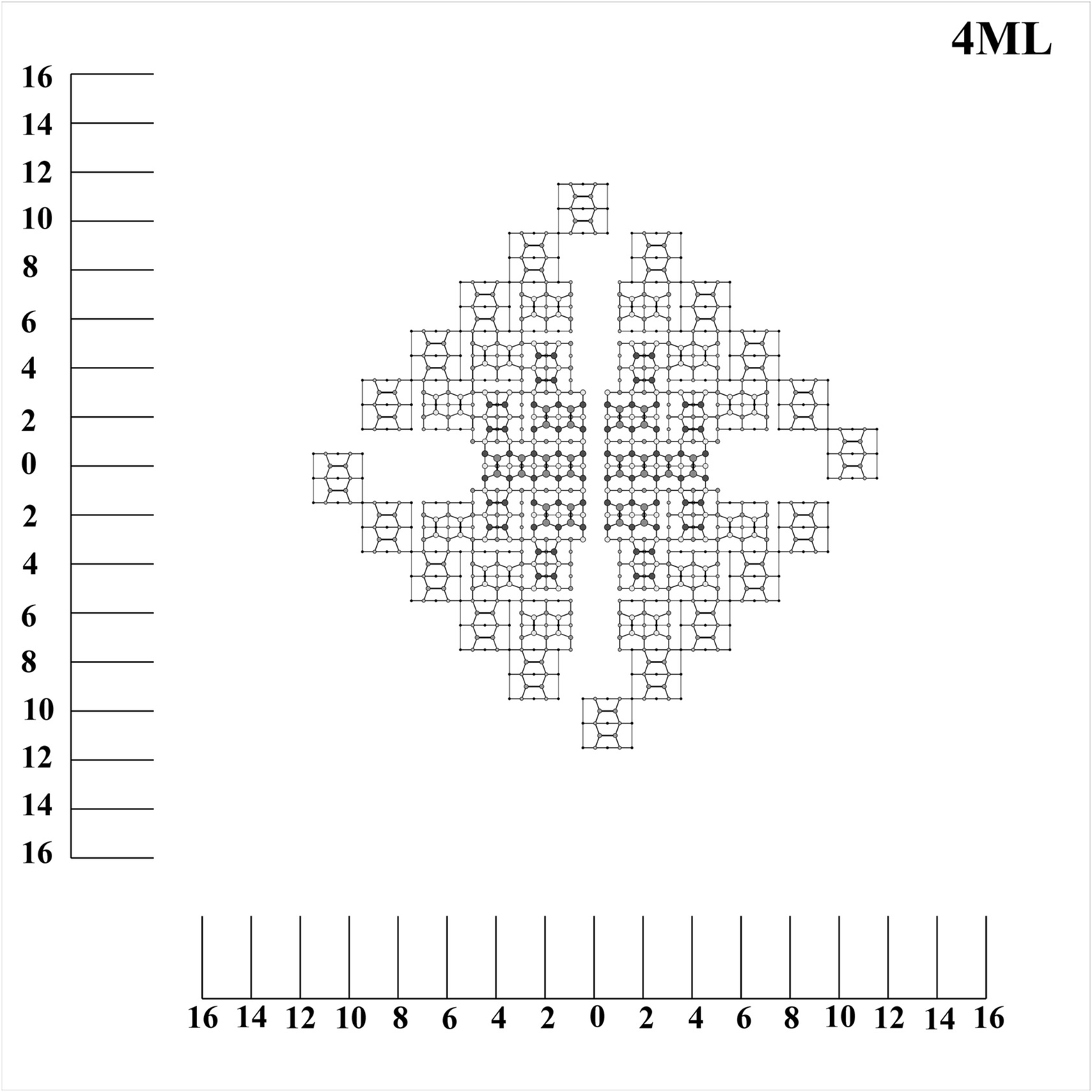}(c)
\includegraphics[scale=.1]{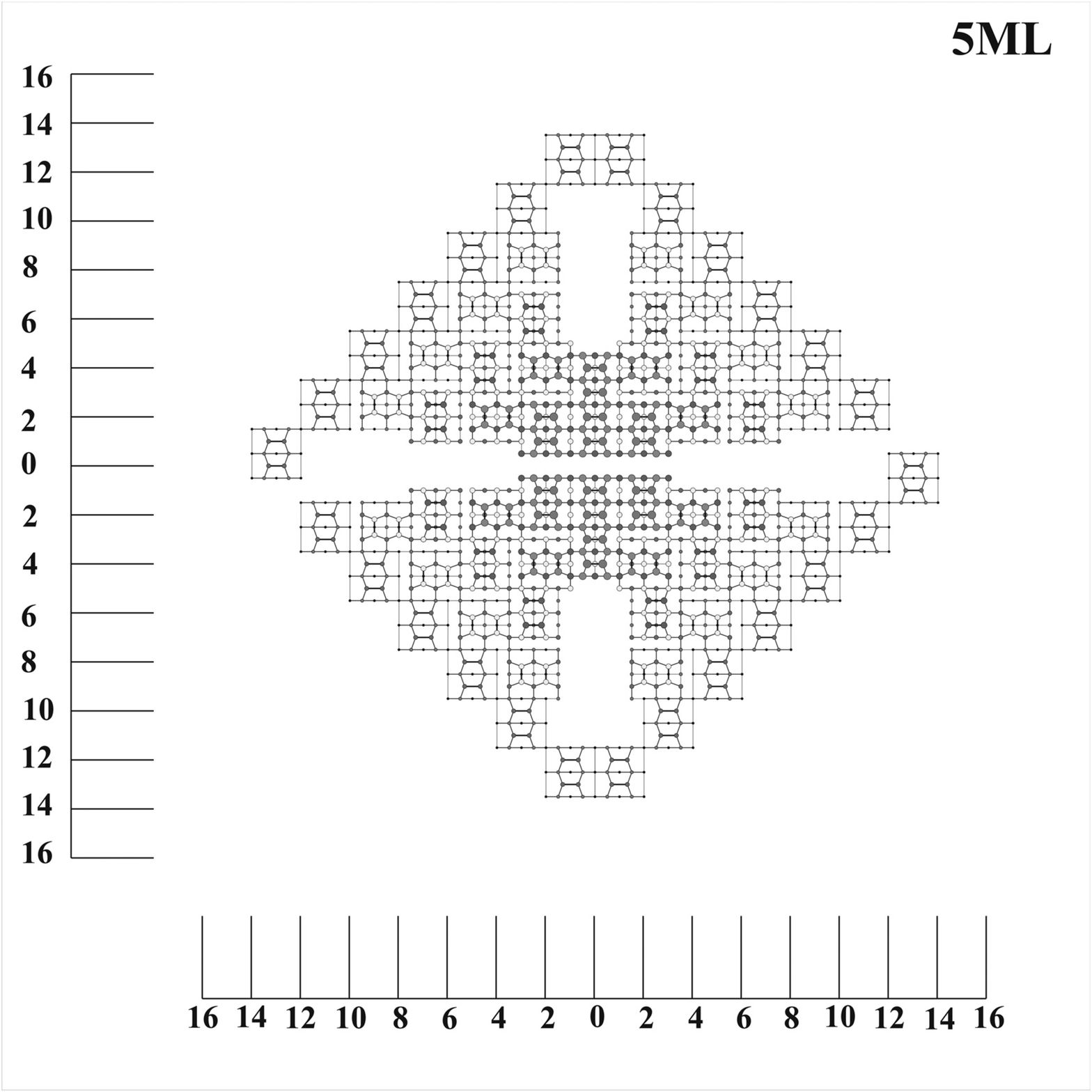}(d)\\
\includegraphics[scale=.1]{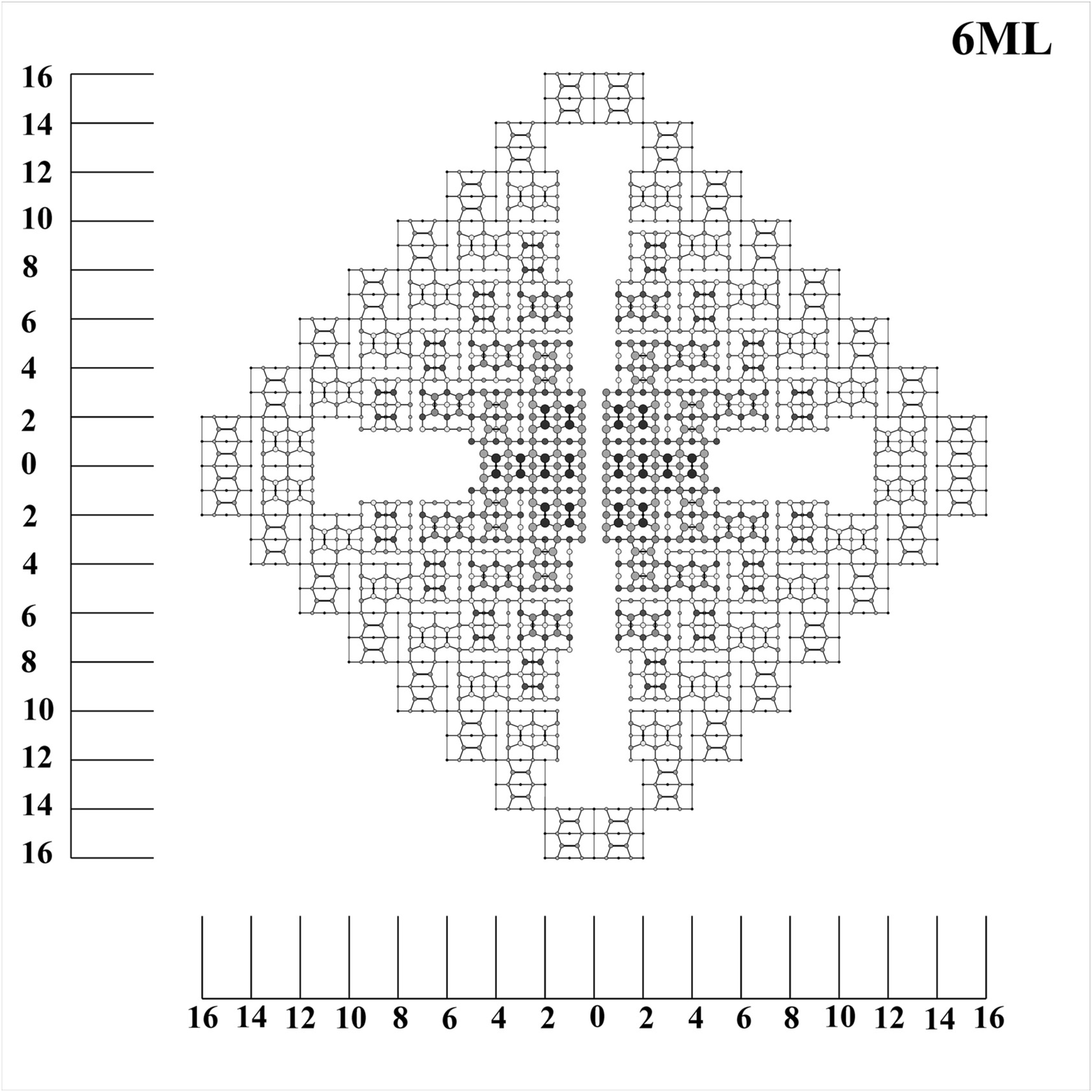}(e)~~~~~~~~~~~~~~~~~~~~~~~~~~~~~~~~~~~~~~~~~~~~~~~~~~~~~~~~~$_{~}$
\caption{\label{fig:non-uniform_pyramid}
A model of non-uniform growth of a pyramidal hut: the cluster height over WL is 2 (a), 3 (b), 4 (c), 5 (d) and 6 (e) ML. Only the structures of the \{105\} facets and the blossom-like apexes are shown. The axes represent the cluster base half-widths along the $<$110$>$ axes expressed in units of elementary translations. The drawings are centered at the (00) point.
}
\end{figure*}

\begin{figure*}
\flushleft
\includegraphics[scale=.24]{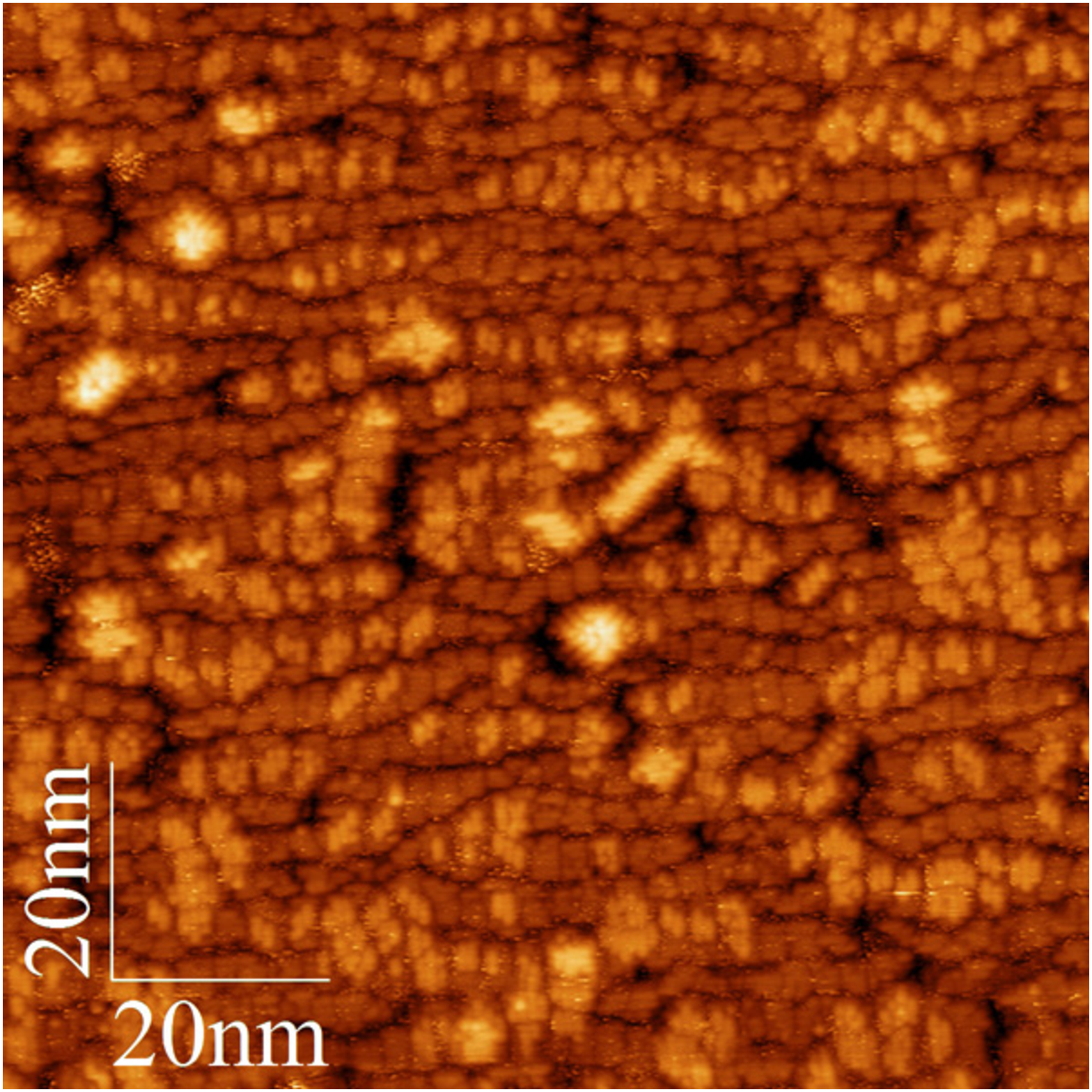}(a)
\includegraphics[scale=.287]{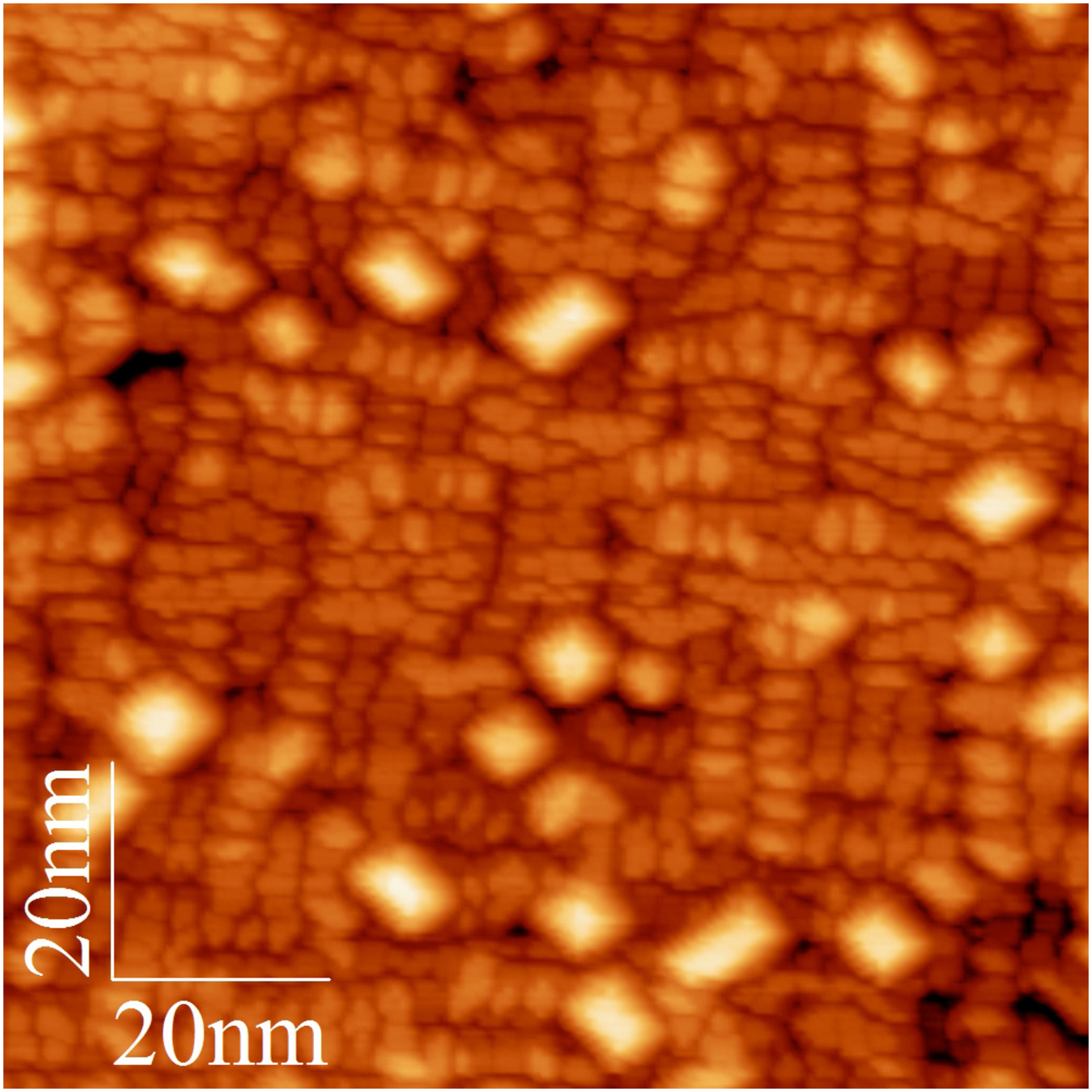}(b)\\
\includegraphics[scale=.18]{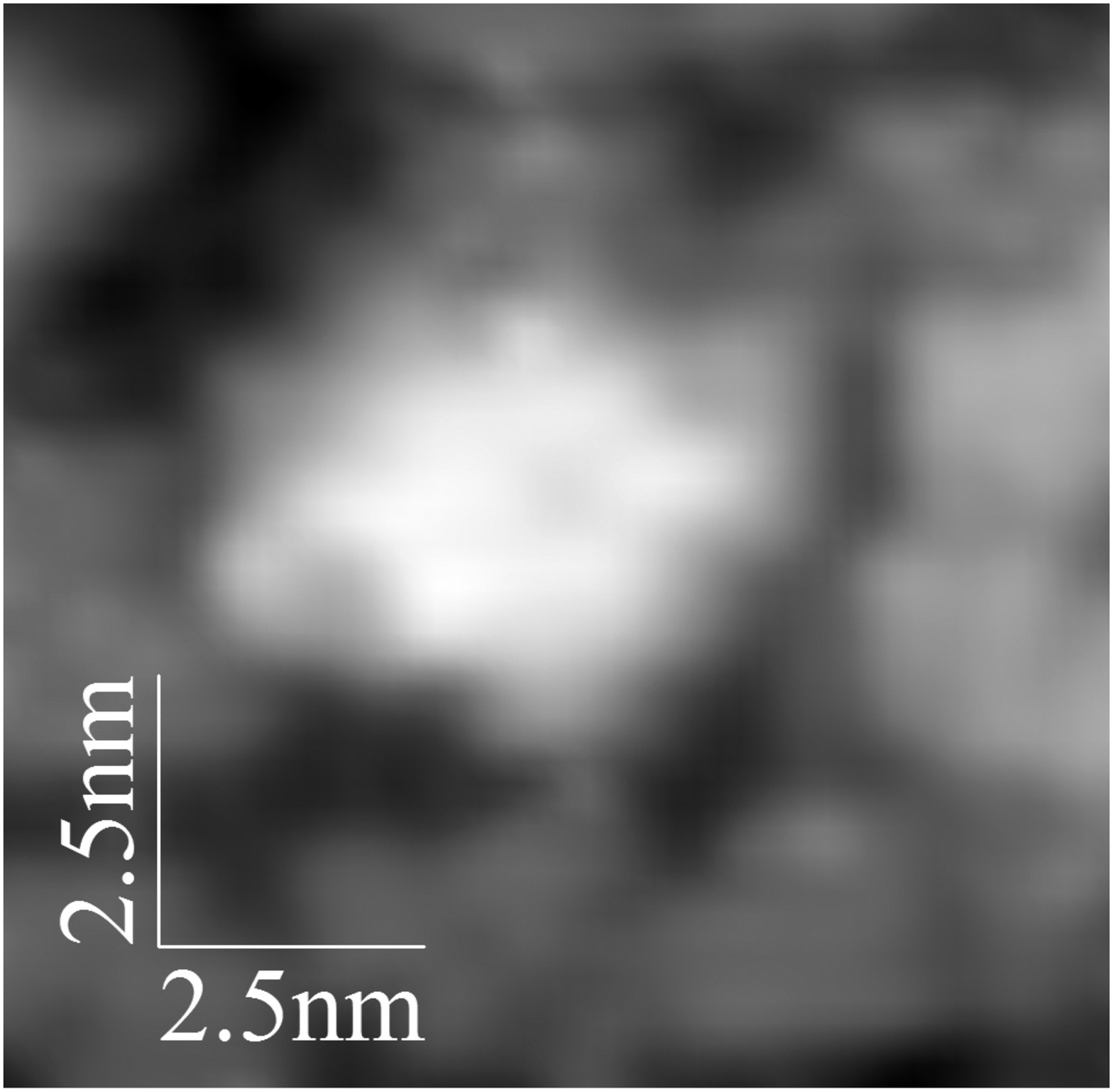}(c)
\includegraphics[scale=.18]{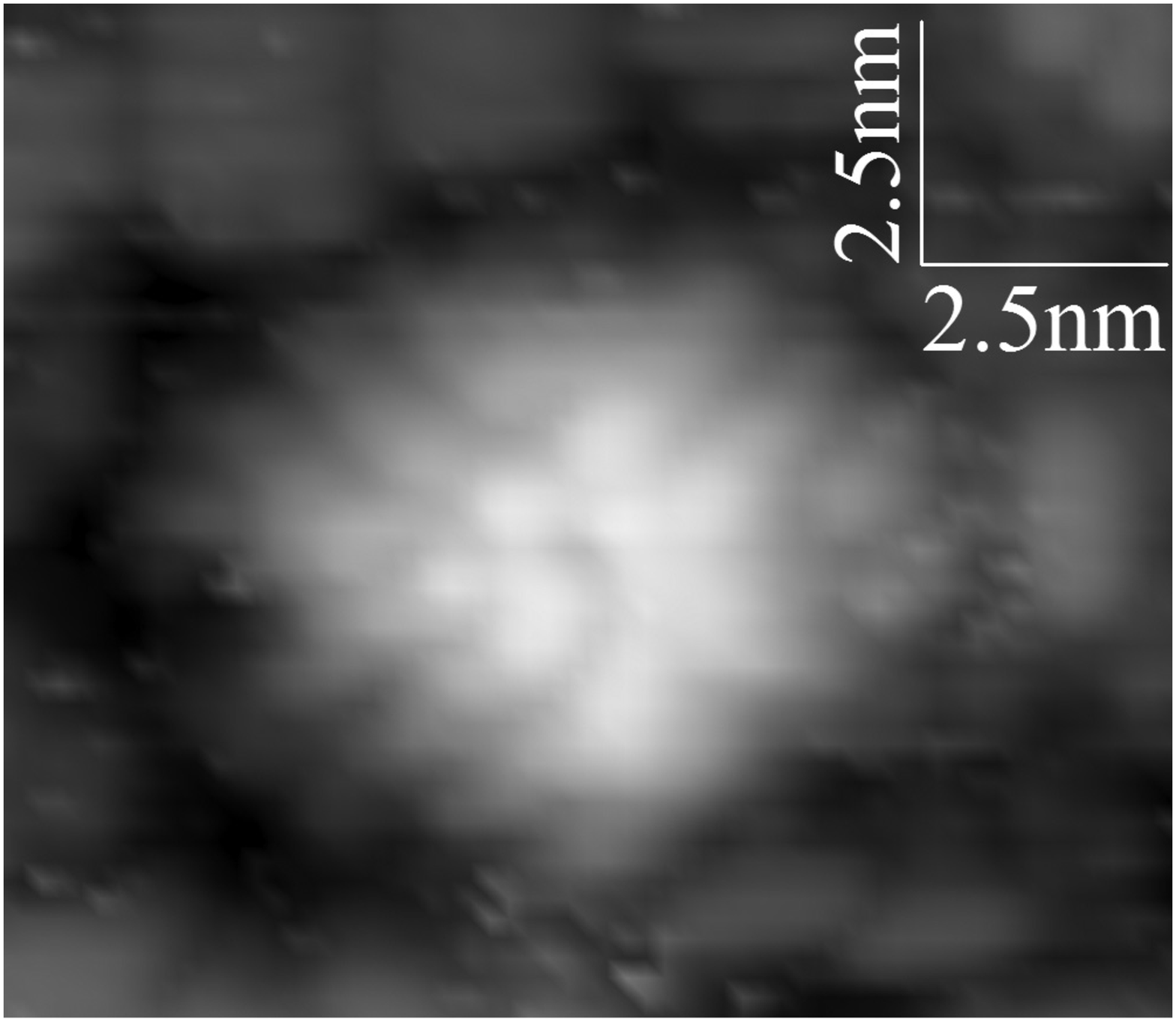}(d)
\includegraphics[scale=.18]{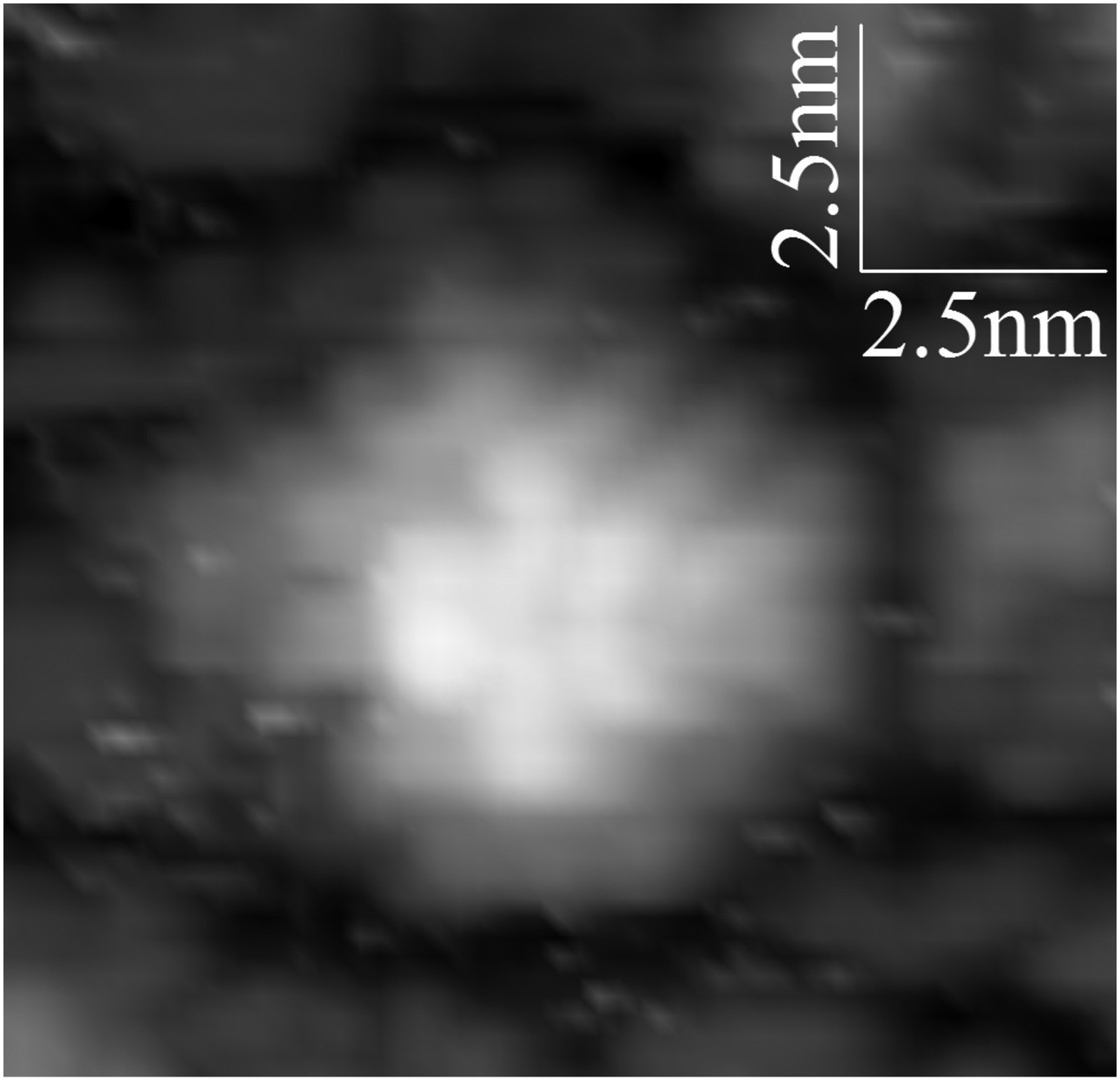}(e)
\includegraphics[scale=.18]{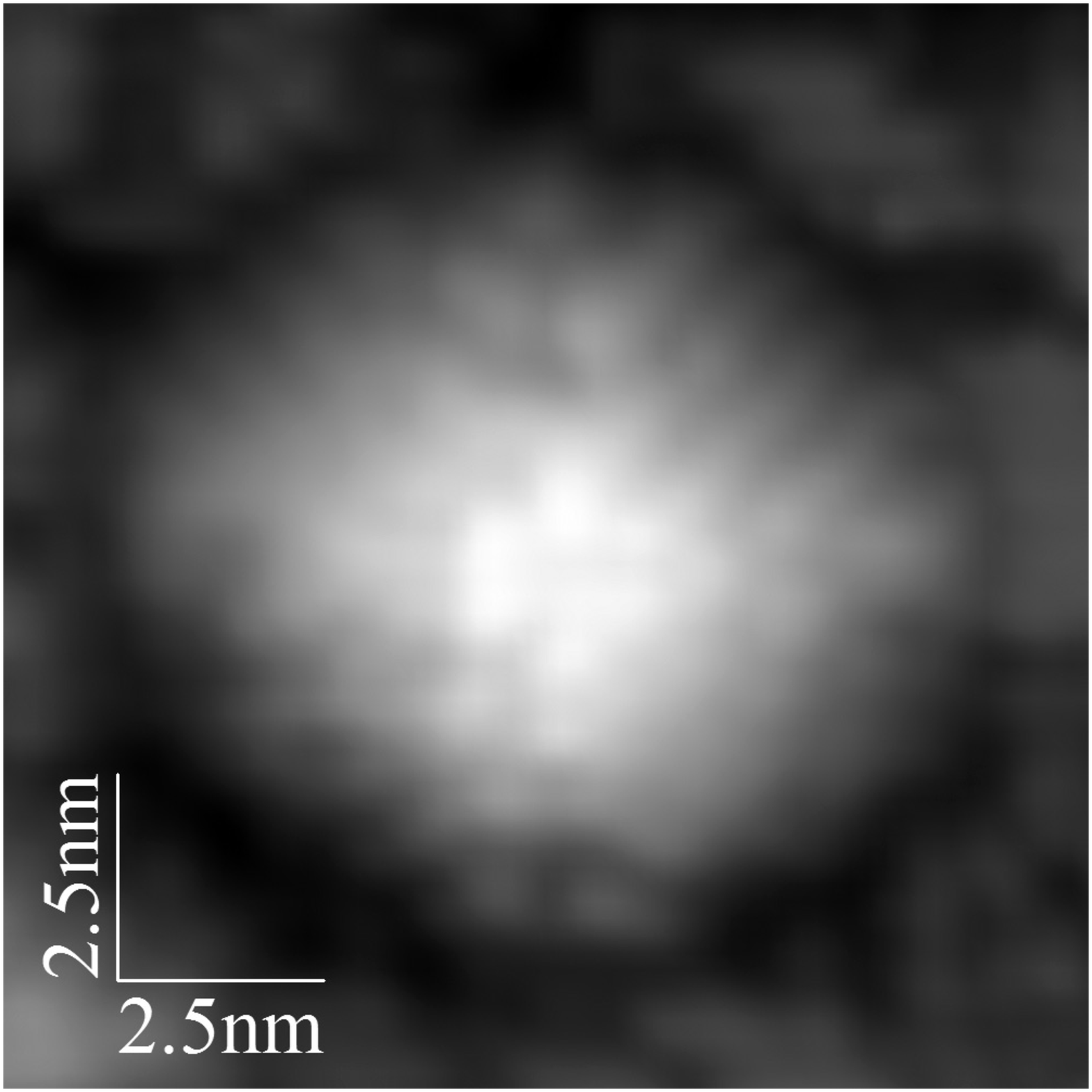}(f)
\includegraphics[scale=1.1]{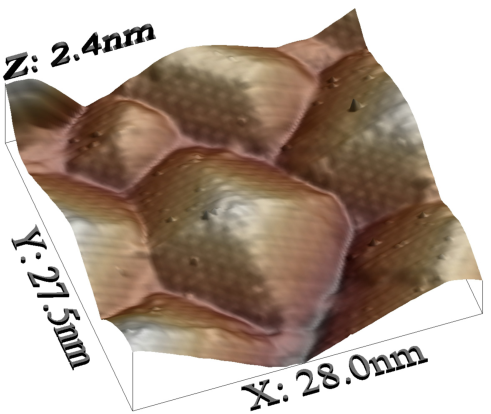}(g)
\caption{\label{fig:array}(Color online) 
STM empty-state images of Ge hut arrays (a),\,(b) and pyramids (c),\,(d),\,(e),\,(f),\,(g) 
grown at 360\textcelsius:
2, 3, 4 and 5-ML pyramids are seen in a variety among wedge-like huts in panels (a) and (b);
micrographs of 2 (c), 3 (d),\,(e) and 5\,ML (f) high (over WL) pyramids  demonstrate fine details corresponding with the diagrams presented in Fig.~\ref{fig:non-uniform_pyramid}, including  the blossom-like structure of a vertex, this structure is also resolved on the top of the mature pyramid presented in image (g) that also corresponds with Fig.~\ref{fig:non-uniform_pyramid};
the Ge coverage is 5.4\,\r{A} in panels (a),\,(c),\,(d) and (e), 6\,\r{A} in (b) and (f), and 10\,\r{A} in (g); 
bias voltage and  tunneling current are
+1.8\,V and 100\,pA in panels (a),\,(c),\,(d) and (e),  
+2.0\,V and 100\,pA in (b), +2.6\,V and 80\,pA in (f), and +2.0\,V and 80\,pA in (g).
}
\end{figure*}

\begin{figure*}
\includegraphics[scale=.4]{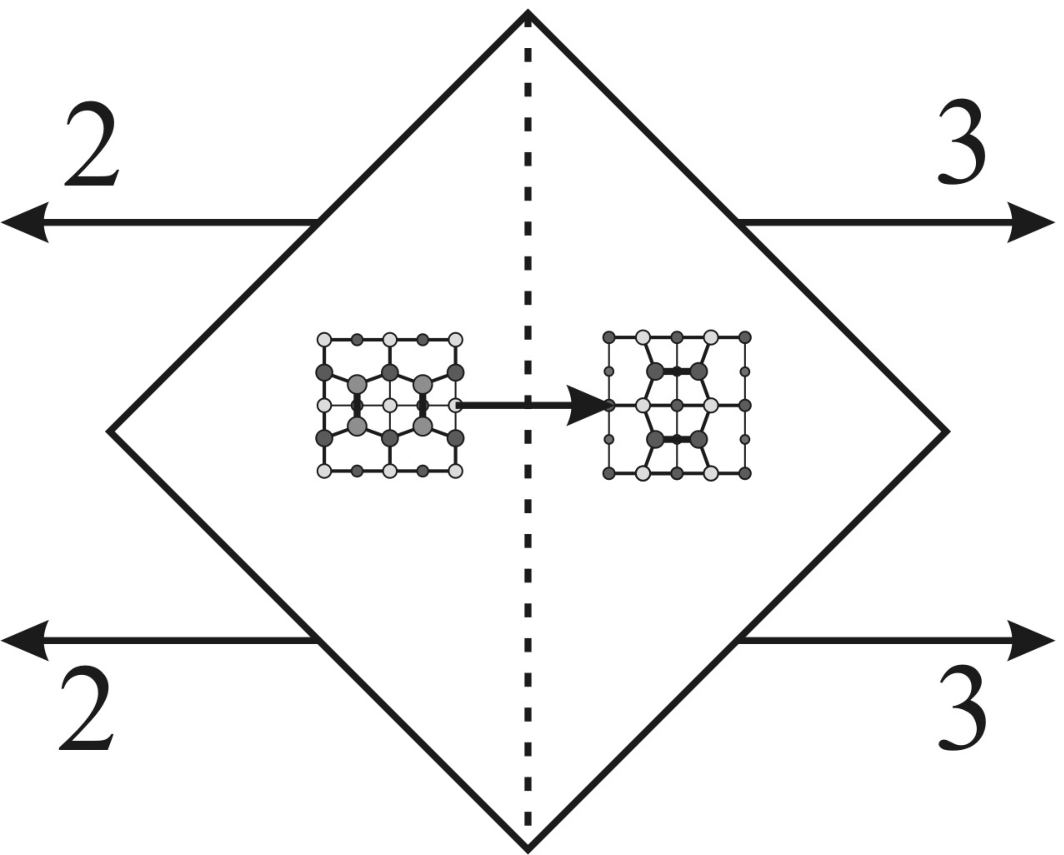}(a)
\includegraphics[scale=.4]{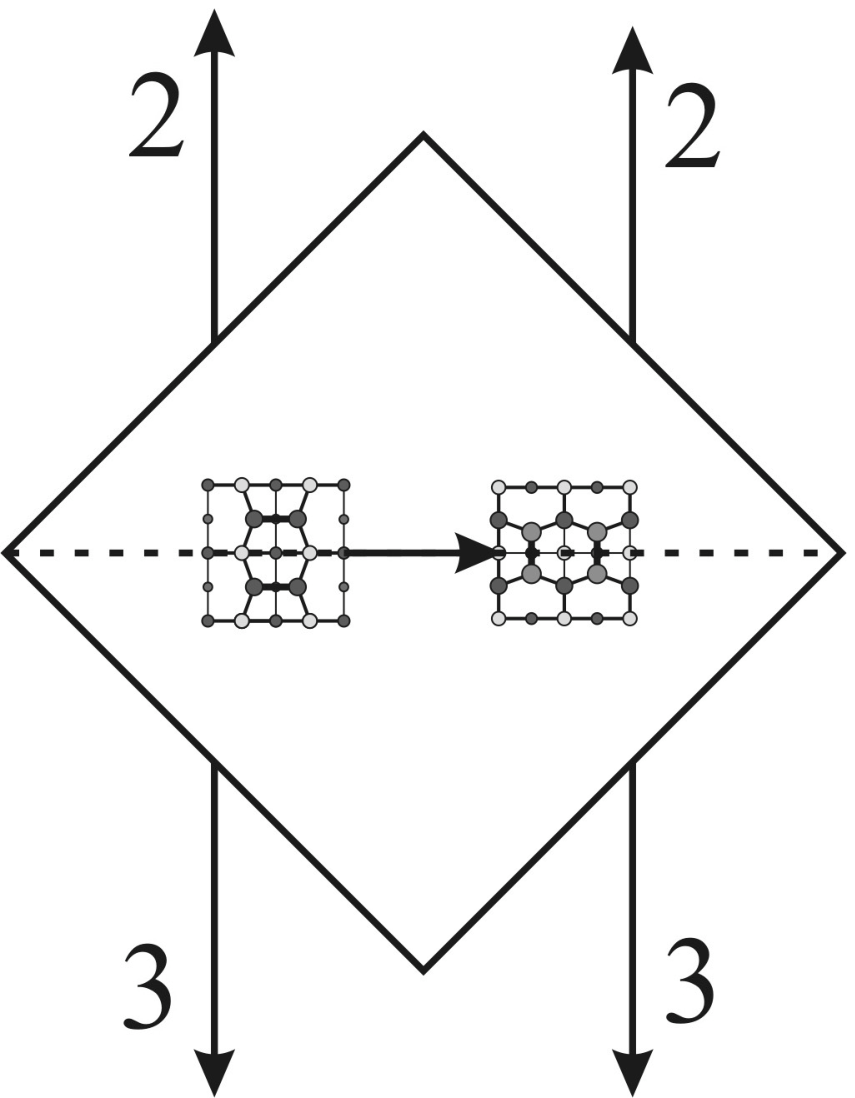}(b)
\includegraphics[scale=.4]{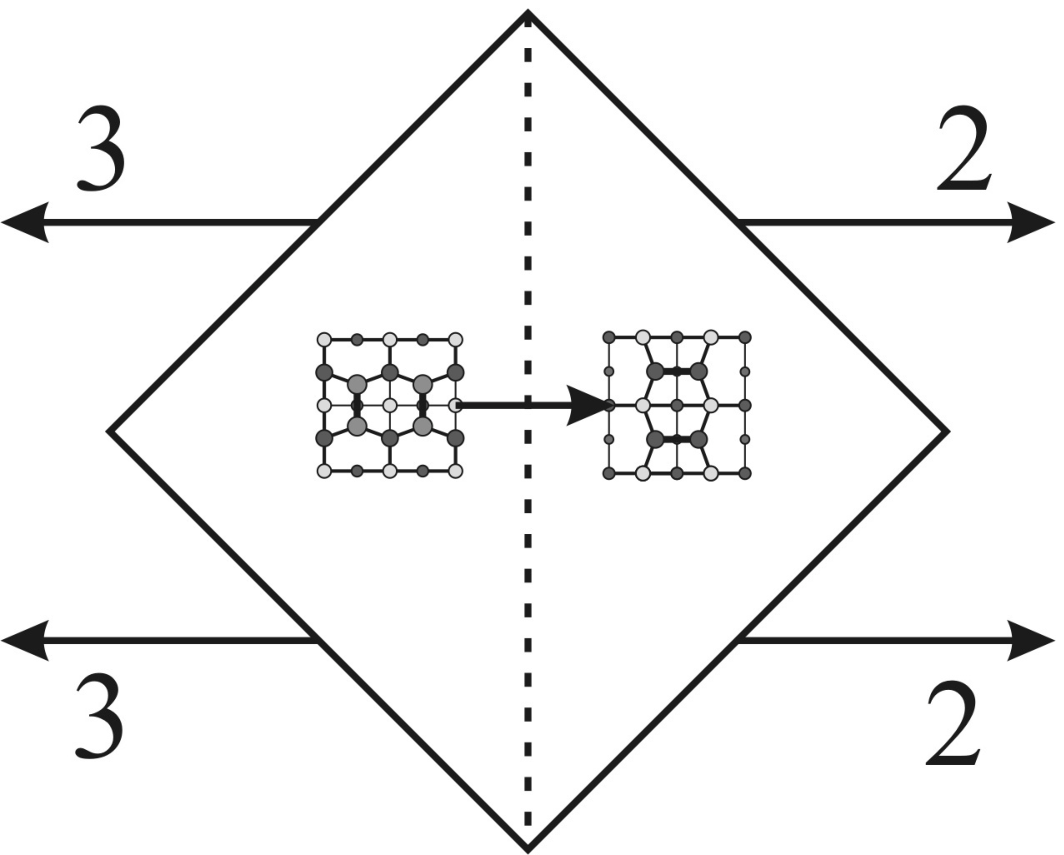}(c)
\includegraphics[scale=.4]{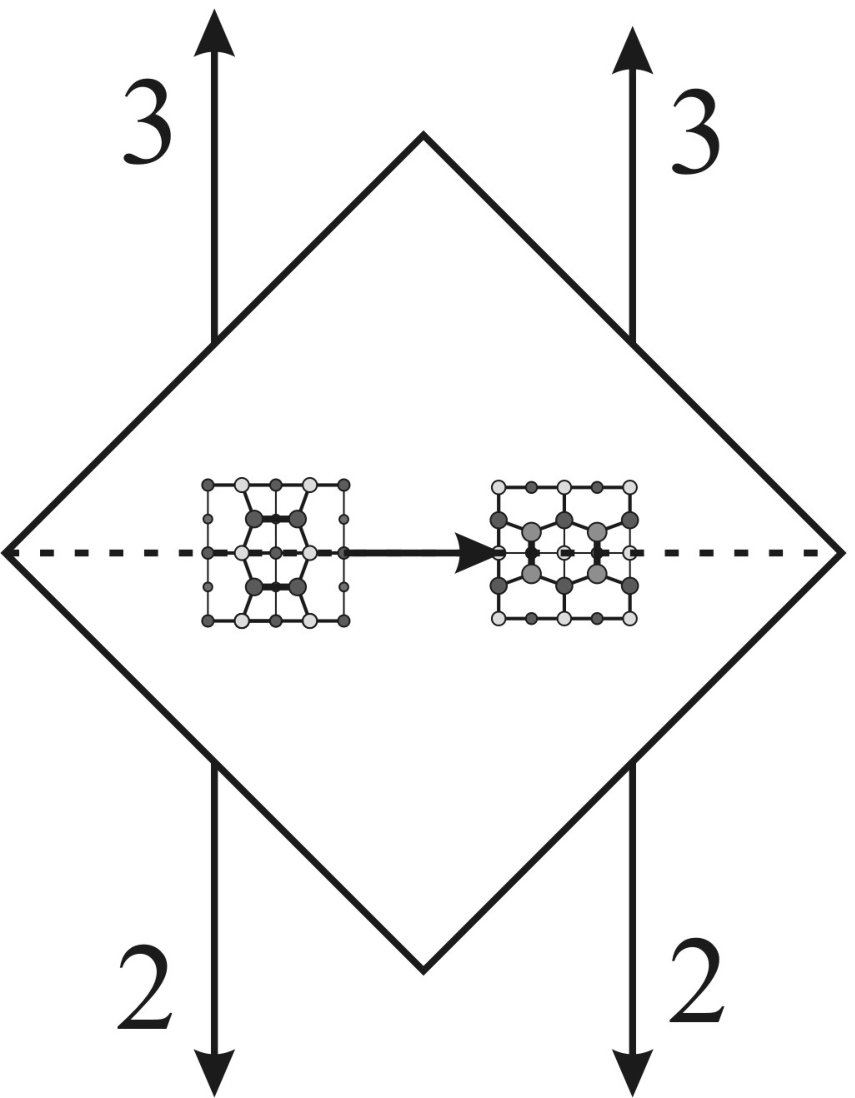}(d)
\caption{\label{fig:pyramid}
Schematic representation of a cycle of the non-uniform growth  of a pyramidal hut-cluster depending on the initial and final direction of dimer rows on its apex.  Each panel correspond to pyramid growth by 1\,ML, e.\,g., to the transitions between growth phases from 2 to 6\,ML shown in Fig.~\ref{fig:non-uniform_pyramid}: 
2\,ML to 3\,ML (a), 
3\,ML to 4\,ML (b), 
4\,ML to 5\,ML (c), 
5\,ML to 6\,ML (d)
(or in the same order of steps from 6 to 10\,ML, from 10 to 14\,ML and so on).
The apex rotation is shown by sketches of the unit cells connected by arrows inside the squares; the arrowheads show the final directions which are vertical in panels (a),\,(c) and horizontal in panels (b),\,(d). Arrows at bases show the $<$110$>$ directions and figures near them display a number of elementary translations to which the corresponding base side is shifted in the indicated direction due to the increase in the cluster height by 1\,ML.
}
\end{figure*}

Figures~\ref{fig:uniform_wedge} to \ref{fig:pyramid} schematically present atomic models of the in-height growth of hut-clusters drawn on the basis of the above experimental data. There are two possible models of processes which, in principal, may describe the growth of huts.\cite{VCIAN2011} The first one implies that the cluster growth goes on due to uniform attachment of Ge atoms to all cluster facets (we refer to this model as \textit{a model of uniform growth}).\cite{VCIAN2011} This means that the width of a cluster in all  $<$110$>$ directions increases by the value of two elementary translations (or in total by four elementary translations along each $<$110$>$ axis) after each step of completion of the cluster facets (Fig.~\ref{fig:uniform_wedge}).\footnote{
Let us remind that the magnitude of an elementary translation along $<$110$>$ axes equals 3.84\,\r{A} for unstrained Si and 4\,\r{A} for unstrained Ge; for a compressively strained Ge film on Si(001), it lies between these values}
 For the simplicity of presentation we do not show the structure of hut edges which is not essential for understanding; we do not consider elongation of wedges along $<$100$>$ axes (the longitudinal growth) either. For the mechanism of uniform growth the structure of a wedge-like hut ridge would depend on  the cluster height\footnote{Usually we measure heights of clusters from the surface of WL although deep fosses,  which can pierce WL and reach a Si substrate (their depth was measured by STM; see, e.g., panel (e) in Fig.~\ref{fig:non-uniform_wedge} and panels (d) to (f) in Fig.~\ref{fig:array}), are sometimes observed near or around bases of formed clusters ($>$\,2\,ML high over WL) at low and moderate coverages when most clusters are not coalesced and free WL surface is broad between huts.\cite{classification,CMOS-compatible-EMRS,VCIAN2011}  Nevertheless measurement of cluster heights from WL is more appropriate to structural investigations than measurement from the substrate surface because clusters nucleate on tops of WL patches and their height from WL better corresponds with their structure and growth processes; in addition it is more suitable for STM studies.}
(Fig.~\ref{fig:uniform_wedge}); the dimer arrangement on the apexes would analogous for the wedge heights of 2\,ML and of 6\,ML with  the base width corresponding to 7\,ML which is shown in Fig.~\ref{fig:uniform_wedge}\,d under the designation of 7\,ML because it would form as a result of 7 rounds of completion of its facets. In other words, the ridge structure of the 2-ML wedge would be reproduced after completion of every next 5\,ML during the cluster growth. In Fig.~\ref{fig:uniform_wedge}, the cluster height of 2\,ML  is chosen as an initial point; the 3-ML cluster is obtained as a result of widening of the base by 2 translations in each of the $<$110$>$ directions. The ridge structures of the 2-ML and 3-ML clusters do not coincide. Further expansion of the cluster is similar. The 6-ML cluster (Fig.~\ref{fig:uniform_wedge}\,c) would have a simple top structure demonstrating \textit{observable ridge narrowing}. After completion of 7 layers the cluster apex would repeat the structure of the 2-ML wedge but the cluster real height would be only 6\,ML over WL,
i.\,e. being 6\,ML in height it would have the base area corresponding to the height of 7\,ML.  According to our observations\cite{classification,VCIAN2011,CMOS-compatible-EMRS} made by STM and to the data obtained by different authors and presented, e.\,g., in Refs.~\onlinecite{Fujikawa_ASS,Iwawaki_SSL,Iwawaki_initial} wedge-like huts of different heights always have the same structure of ridges (which is clearly seen in Fig.~\ref{fig:non-uniform_wedge}\,e). So, we can conclude that the above model of uniform growth of wedges finds no confirmation by experimental data in the case of Ge huts.

The other model 
implies that  huts grow due to non-uniform attachment of Ge atoms to each of the \{105\} facets (we refer to this model as  \textit{a model of non-uniform growth});\cite{VCIAN2011} this means that the cluster width in each of the $<$110$>$ directions increases by two or three translations when the cluster height rises by 1\,ML. Such process of growth ensures the independence of atom arrangement on the hut apex of the cluster height, as observed in experiments. This model for the case of wedge-like huts is illustrated by Fig.~\ref{fig:non-uniform_wedge}. The direction in which the wedge base expands to 3 translations changes with the increase in the cluster height by 1\,ML because this direction is  determined by the direction along which the rows run forming the topmost (001) terrace. This is schematically presented in Fig.~\ref{fig:wedge}. Arrows in the drawing show the $<$110$>$ directions  along which Ge atoms attach to the corresponding facet; the figures next to the arrows display a number of translations to which the facet shifts in this direction. The hut cluster is represented by the rectangle divided to two sections by the diagonal. On one side from the diagonal the cluster base expands to two translations in each of the indicated $<$110$>$ directions; on the other side the growth process is more complicated. The direction of the base expansion to three translations is determined as follows (Fig.~\ref{fig:wedge}):
If a cluster, which originally had, e.\,g., horizontally oriented  rows on its ridge (Fig.~\ref{fig:wedge}\,a), increases its height by 1\,ML and its apex rows become vertically oriented  (Fig.~\ref{fig:wedge}\,a)  then a direction of  its base expansion to tree translations coincides with the resultant direction of dimer rows on its top.  The adjacent facet situated on the same side of the cluster with respect to the diagonal also grows following a more complex rule than the facets situated on the opposite side: its base expands in two directions at once but to different number of elementary translations---expansion to one translation goes on in the resultant direction of the dimer rows on  the ridge whereas a direction of the base side shift by two translations coincide with the initial direction of rows on the cluster top (Fig.~\ref{fig:wedge}\,a). The analogous process for the next step of completion of the cluster facets and cluster growth by the next 1 ML is presented in Fig.~\ref{fig:wedge}\,b.

Fig.~\ref{fig:non-uniform_pyramid} demonstrates schematics of growth of pyramidal clusters. According the STM observations atom configuration on pyramid vertices does not depend on the cluster height and coincide with that in the pyramid nucleus\cite{Hut_nucleation,classification,CMOS-compatible-EMRS} (compare Fig.~\ref{fig:nuclei}\,(a),\,(b) and Fig.~\ref{fig:array}\,(c) to (g))  so,  like in the former case, the process the pyramid growth corresponds with the  model of non-uniform growth. Fig.~\ref{fig:pyramid} illustrates this process. During the pyramid growth by 1\,ML its base sides move either by two or by three translations in the directions shown by arrows in Fig.~\ref{fig:pyramid}. The direction of the base expansion to three translations is determined as follows (Fig.~\ref{fig:pyramid}\,a): as contrast to the case of the wedge-like cluster, this direction is normal to the final direction of  dimer rows on the cluster vertex; two facet bases on one side from the diagonal (which is shown by the dashed line in Fig.~\ref{fig:pyramid}\,a) move by tree translations in the indicated direction whereas the rest two facet bases situated on the other side from the diagonal are shifted by only two translations in the opposite direction. Let us consider the initial phase of the pyramid growth depicted in  Fig.~\ref{fig:non-uniform_pyramid}. A diagram shown in Fig.~\ref{fig:pyramid}\,a corresponds to the transition from 2\,ML to 3\,ML in the pyramid height (Fig.~\ref{fig:non-uniform_pyramid}\,a,\,b); as a result of this step a symmetrical 2-ML pyramid transforms into slightly asymmetrical 3-ML one because of the process of non-uniform growth. The next transition between 3 and 4\,ML (Fig.~\ref{fig:non-uniform_pyramid}\,b,\,c) happens in accordance with the process plotted in Fig.~\ref{fig:pyramid}\,b which also results in formation of a slightly asymmetrical 4-ML pyramid with the apex structure rotated 90{\textdegree} with respect to the previous one; the process diagram is also seen to be rotated 90{\textdegree} clockwise. The next  rotation of 90{\textdegree}  clockwise of the cluster expansion process (Fig.~\ref{fig:pyramid}\,c), corresponding to the next step of completion of the cluster facets, forms a 5-ML pyramid with the tiny violation of its symmetry (Fig.~\ref{fig:non-uniform_pyramid}\,c,\,d). And finally, a symmetrical 6-ML high pyramid forms (Fig.~\ref{fig:non-uniform_pyramid}\,d,\,e) as a result of the step schematically drown in Fig.~\ref{fig:pyramid}\,d which is again rotated 90{\textdegree} clockwise; this cluster restores the symmetry of the 2-ML one. (Notice that all the above mentioned violations of the pyramid symmetry result in   difference in the length of base diagonals of only 1 elementary translation.) Afterwards the cycle repeats resulting in formation of higher pyramids. 
(STM micrographs of 2, 3 and 5-ML pyramids illustrate this model in Fig.~\ref{fig:array}\,(c) to (f)).

Thus, a cycle of four consecutive steps (processes) of the pyramid expansion (completion), each rotating 90{\textdegree} with respect to the previous one, and all rotating in the same direction---clockwise or anticlockwise,  describes  a complete process of the pyramid growth; and only every fourth step in the row (starting from the 2-ML one) results in appearance of the symmetrical cluster (the 6-ML one, 10-ML one, 14-ML one, etc). The pyramids of different heights are a little asymmetrical. This maybe means that only the symmetrical pyramids are stable while the rest are metastable.

This hypothesis would probably allow us to explain the decay of the pyramid number density and their relative fraction in the  hut arrays observed during the low-temperature growth which eventually results in their virtual vanishing from the arrays at high Ge coverages.\cite{classification, Hut_nucleation, VCIAN-2012} The explanation might be as follows: Pyramids of the first stable height (2 ML) are often observed at low Ge coverages (less or around 6\,\AA);\cite{initial_phase, CMOS-compatible-EMRS} numerous 3, 4 and 5-ML pyramids are also observed at Ge coverages some less or about 6\,\r{A} (Fig.~\ref{fig:array}) when the clusters are small enough and the distances between them are large in comparison with their dimensions.\cite{initial_phase, CMOS-compatible-EMRS,Hut_nucleation,VCIAN2011} In these conditions, if a flux of Ge atoms arriving on the surface is sufficient to feed all the growing huts and a competition between huts for Ge is actually absent, all pyramids, both stable and metastable, obtain enough material to complete their facets and grow remaining in the array. Pyramids of 6\,ML high are stable but they form at greater coverages, from 6 to 8\,\AA, when all huts become bigger, more actively consume Ge and gaps between then become smaller.\cite{classification,VCIAN2011} Only stable pyramids can survive among aggressively  growing competitors,  and metastable ones likely loose their substance in favor of stable counterparts (predominantly wedges), probably until decreasing in height reach 2\,ML. These little pyramids are then easily overgrown by large huts. Some pyramids, probably those which nucleated at early stages and had enough time to grow, succeed to reach the stable heights of 10, 14 or 18\,ML over WL being sufficiently large to successfully compete with other clusters even at intermediate  heights---between 6 and 10\,ML, 10 and 14\,ML, etc.---when the effect of minor asymmetry is negligible and they become virtually stable; these pyramids remain in arrays  and are observed as residual fraction in amount of around 10\,\% of the total  number density of huts or as array defects.\cite{Hut_nucleation, CMOS-compatible-EMRS, classification, VCIAN-2012,defects_ICDS-25} 

Notice also, that at high temperatures,  when the cluster number density is small and the Ge dimer mobility on WL is high, pyramids become more stable than wedges which for some reason do not nucleate in the arrays.\cite{Nucleation_high-temperatures} Pyramids can grow even being metastable at intermediate heights in these conditions as they do not experience rivalry for Ge with their counterparts.

\subsection{Nucleation of pyramids and wedges: A scenario of a common embryo}

\begin{figure}
\includegraphics[scale=.8]{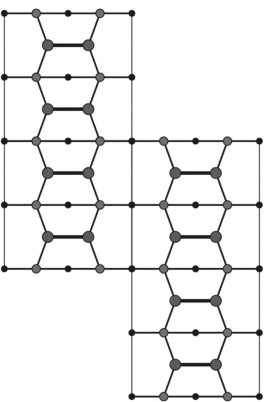}(a)
\includegraphics[scale=.28]{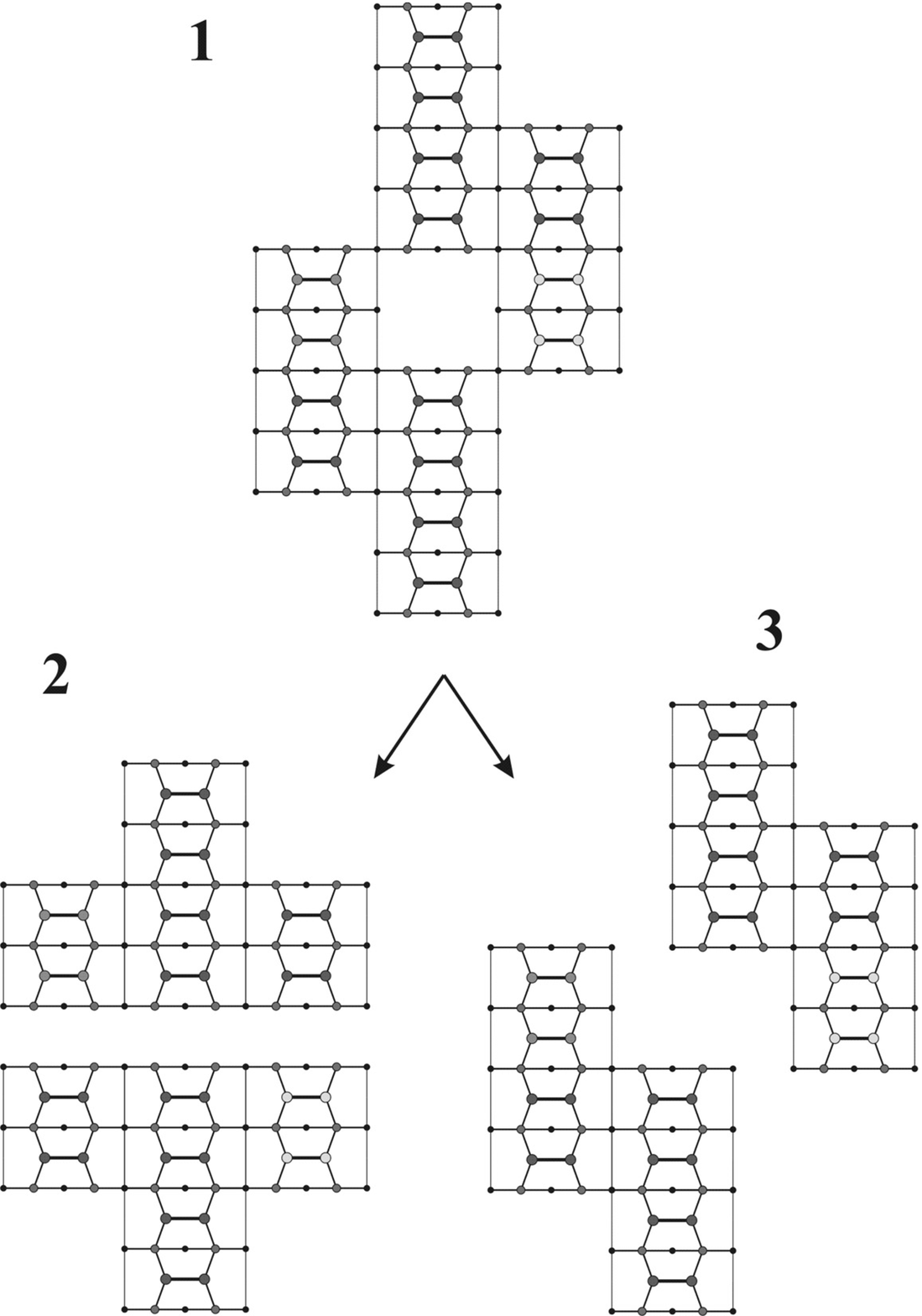}(b)
\caption{\label{fig:nucleation}
Schematics of hypothetic processes of pyramidal and wedge-like hut nuclei formation from a single embryo: the embryos, like that shown in panel (a), form the nuclei (b) by pairing into a 16-dimer formation (1) followed by reconstruction into the blossom-like nucleus of a pyramid (2)  or into the structure (3) preceding the formation of the wedge unit  presented in Fig.~\ref{fig:reconstruction}\,d. 
}
\end{figure}

Presently, hut nucleation poses a lot of questions, and the issue of appearance of two species of nuclei on tops of WL patches to relieve the strain,\cite{Hut_nucleation} instead of single one which seems to be quite enough, is the most intriguing of them.\cite{Nucleation_high-temperatures,initial_phase} Now, we propose a scenario which could partially explain this phenomenon by reducing the quantity of initial structures and deriving both nuclei from a common precursor (``embryo''). 
Fig.~\ref{fig:nucleation} illustrates this  scenario: We suppose that each embryo consists of two rows of dimes; and each row is composed by four dimers (Fig.~\ref{fig:nucleation}\,a). Such structures are abundant on WL at low coverages (frm 5 to 6\,ML)\cite{Nucleation_high-temperatures, initial_phase, VCIAN2011, CMOS-compatible-EMRS} and can be easily recognized in the STM images (Fig.~\ref{fig:array}\,a; they also can be seen in the images presented in Ref.~\onlinecite{Iwawaki_initial}).

Connection of two such embryos into a single whole leads to formation of a structure that has been interpreted by us as a nucleus of wedges\cite{Hut_nucleation} (Fig.~\ref{fig:nucleation}\,b, figure 1). Then this formation can stabilize by reconstruction into the structure of the nucleus  of pyramids with all bases sides aligned with $<$100$>$ which is ready to formation of all four \{105\} facets. Such structures are seen in STM images at initial phases of hut array formation at low Ge coverages (Fig.~\ref{fig:array}\,a,\,b) and temperatures.\cite{Hut_nucleation, Nucleation_high-temperatures, initial_phase, VCIAN2011, CMOS-compatible-EMRS} A structure denoted by figure 3 in Fig.~\ref{fig:nucleation}\,b, to which the formation (1) can also transform, likely is not stable: it has only two base sides running along  $<$100$>$ and cannot form four \{105\} facets. This structure is  observed in STM micrographs very rarely. Probably, namely this explains the fact that height of wedge-like huts usually is greater than 2\,ML.  Formation of  2\,ML high stable wedge-like  huts likely  occurs rapidly through a phase transition, which is a shift in one of the cluster parts by half translation (Fig.~\ref{fig:nucleation}\,b, transition from the state 1 to the state 3), followed by attachment, as discussed above, of  specifically arranged Ge adatoms (ad-dimers) of the second monolayer.\cite{Hut_nucleation}

\subsection{\label{sec:WL_structure}Wetting layer structure  before hut array appearance}

\begin{figure*}
\includegraphics[scale=1]{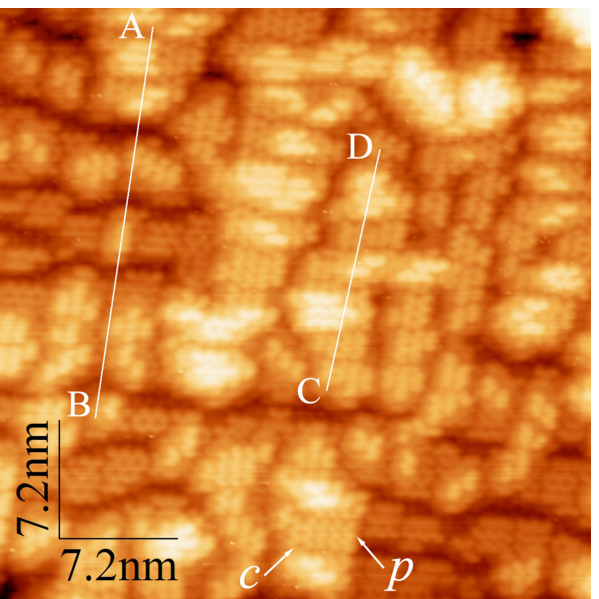}(a)
\includegraphics[scale=.25]{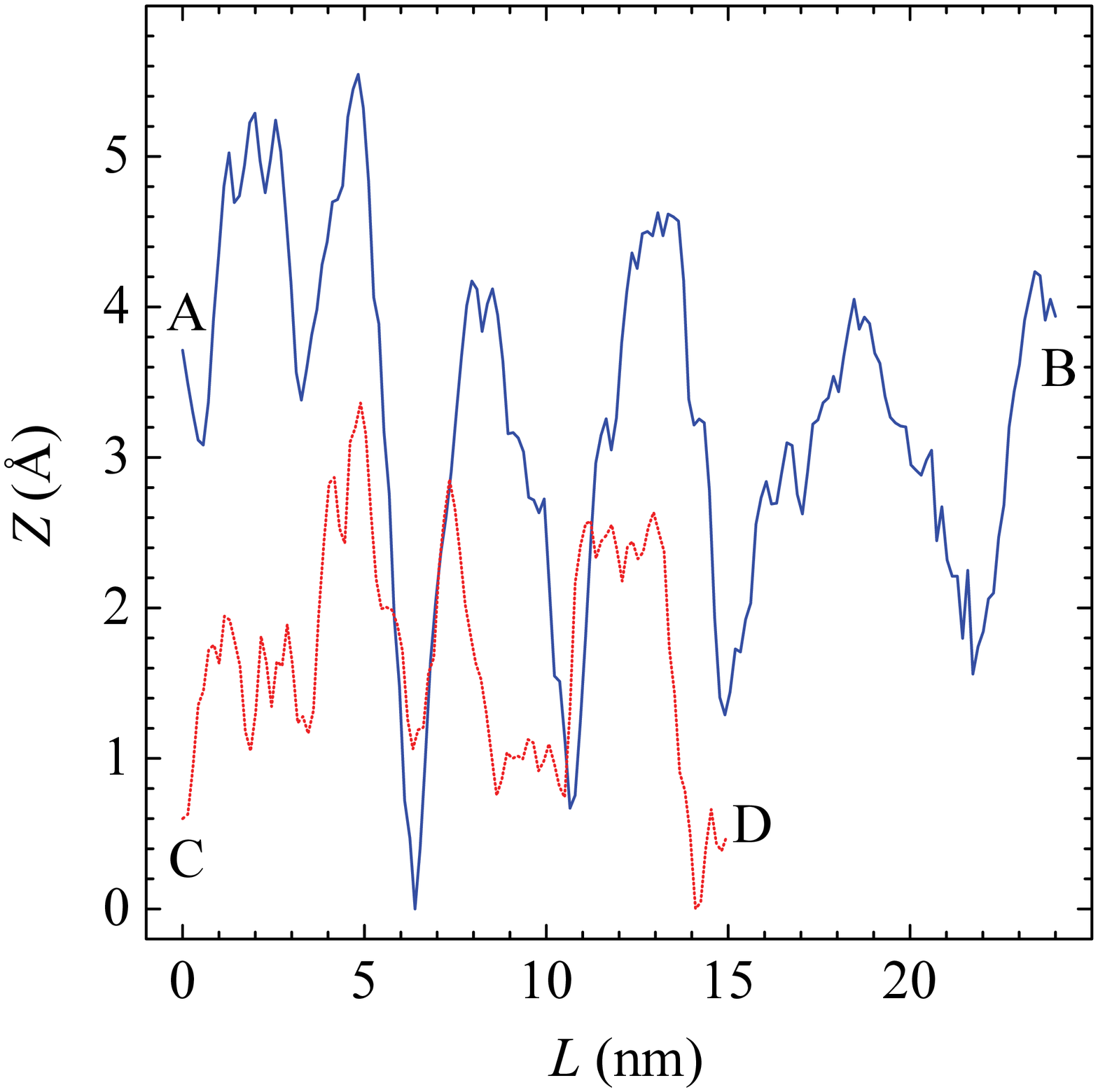}(b)
\includegraphics[scale=1.3]{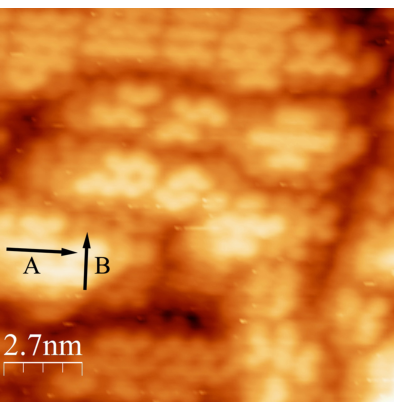}(c)
\caption{\label{fig:WL_structure}(Color online) 
STM fiiled state image [$-$1.992\,V, 100\,pA] of Ge WL (a) deposited  at 360{\textcelsius} ($h_{\rm Ge}= 4$\,\AA) and surface roughness profiles (b) taken along AB and CD lines; letters $c$ and $p$ indicate $c(4\times 2)$ and $p(2\times 2)$ reconstructions, respectively; (c) the same as in panel (a) at higher magnification [$-$1.992\,V, 150\,pA].
}
\end{figure*}

As we have already mentioned above,\cite{Note1} WL patches are isolated from one another. It can be quite definitely concluded from Fig.\,\ref{fig:WL_structure}\,a,\,b. in which deep trenches of row vacancies are seen to slot whole WL often reaching the Si substrate (see the AB line). trenches of vacancy rows are not so deep: their depth is only one ML (or sometimes 2\,ML if measured from the topmost rows, see the CD line). However, $c(4\times 2)$ and $p(2\times 2)$ reconstructions can coexist on adjacent patches and even on one and the same patch as it is shown by arrows in the panel (a) (this is an important distinction of WL formed at low temperatures from that formed at $T>600${\textcelsius} when only $c(4\times 2)$ forms\cite{Nucleation_high-temperatures}).
\footnote{Reflected high-energy electron diffraction demonstrates that at low temperatures Ge clusters nucleate on $(2\times 1)$-reconstructed WL surface;\cite{Nucleation_high-temperatures, VCIAN-2012} this is not the case for the high-temperature growth during which WL remains unreconstructed until the temperature decreases below 600{\textcelsius} after the Ge deposition}
So, we can suppose that WL patches can efficiently relieve the strain and the lattice parameter of their top layers can return to that of the unstrained Ge before huts start to nucleate. 
Previously, the authors of Ref.~\onlinecite{Nikiforov_Ge-Si_RHEED, *Nikiforov_Ge-Si_RHEED-oscillations} observed such relaxation (and even some tensile strain just before the onset of hut nucleation) {\it in situ} by the recording diffractometry at Ge deposition temperatures.\cite{[Analyzing possible mechanisms of Ge hut nucleation Teys  has also come to the same conclusion concerning strain relaxation in WL{; }]Teys_WL}

We have tried to verify those observations by STM  using images taken at room temperature ($T_{\rm gr}=360$\textcelsius, $h_{\rm Ge}=4$\,\AA, quenching during cooling) at high magnification with atomic resolution (like that presented in Fig.\,\ref{fig:WL_structure}\,c) plotting roughness profiles and measuring mean values of interatomic distances along (arrow A) and across (arrow B) dimer rows on different terraces of WL patches. We have also analyzed Fourier transforms of these micrographs. 
We failed to reveal the stain relaxation along the rows on the second and the third ML of the WL patches (the first ML was too deep and its area was too small to reliably measure distances)  when the rows were crossed by  shallow and narrow vacancy-row trenches. We obtained the lattice parameter $a_{\|[110]{\rm A}}\approx 3.8\,{\mathrm \AA}=a_{\|[110]{\rm Si}}$ in these rows. If the rows had enough room to elongate, e.\,g., if they ended  by the deep and wide row-vacancy trenches at least on one side or if the were short segments of rows on  tops of patches they turned out to be relaxed; $a_{\|[110]{\rm A}}\approx 4\,{\mathrm \AA}=a_{\|[110]{\rm Ge}}$ (the measured values sometimes reached 4.2\,\AA).

The lattice constant measured across the rows turned out to be $a_{\|[110]{\rm B}}\approx 4\,{\mathrm \AA}=a_{\|[110]{\rm Ge}}$ or even some greater for the third and forth ML (deeper terraces appears to remain strained but the values of  $a_{\|[110]{\rm B}}\approx 3.8\,{\mathrm \AA}=a_{\|[110]{\rm Si}}$ measured on them does not seem to be quite reliable). We explain this observation by the effect of row-vacancy trenches which bound the WL patches on both sides in this direction.

Our data are seen to be in very good agreement with the previous results of Ref.~\onlinecite{Nikiforov_Ge-Si_RHEED, *Nikiforov_Ge-Si_RHEED-oscillations}. 
This allows us to conclude that huts nucleate on relaxed top layers of isolated WL patches rather than on a compressively strained patched Ge film. They turned out to appear not to relax stress in WL but because further planar growth would give rise to a new increase in strain in the WL patches.
This circumstance should be taken into the account in numerical simulations and development of theories of Ge/Si(001) hut nucleation.

\section{Conclusion}

In conclusion of the article, we would like to emphasize its main statements.

On the basis of data of recent STM investigations of nucleation and growth of Ge huts on the Si(001) surface  in the process of molecular beam epitaxy, we have proposed structural models of growing Ge/Si(001) pyramids and wedges and found the huts regardless of their shapes to grow non-uniformly, expanding their bases by different number of translations in different $<$110$>$ directions as a result of increase in their height by 1\,ML,  in order to conserve the dimer arrangement on their apexes, as observed experimentally. We have concluded from the model of non-uniform growth that growing pyramids, starting from the second monolayer, increase their heights via cyclic (recurrent) formation of slightly asymmetrical and symmetrical shapes, with symmetrical ones appearing after addition of every  fourth (001) monolayer. We suppose that only symmetrical configurations of pyramids composed by 2, 6, 10, 14, etc. monolayers over WL are stable. This might explain less stability of pyramids in comparison with wedges in dense arrays obtained at low Ge deposition temperatures.

We have proposed and discussed possible processes of nucleation  of pyramids and wedges on WL patches from the same embryos composed by 8 dimers grouped in two rows through formation of 1\,ML high 16-dimer nuclei different only in the symmetry of arrangement of their dimers. The proposed models, which  consider the very beginning of formation of Stransky-Krastanov Ge/Si(001) clusters on atomic level, seem to show the way on which the issue of simultaneous nucleation, with equal likelihoods, of two species of huts at low temperatures of Ge deposition  can be solved.

And finally, we conclude from precise STM measurements that huts nucleate on relaxed top layers of isolated WL patches to prevent an increase in strain in WL patches  because of accommodation of excess Ge on their tops.

\begin{acknowledgments}

This research has been financed by the Ministry of Education and Science of Russian Federation through the grants No.~8744 and 14.132.21.1395; it has also been supported by the Russian Foundation for Basic Research through the grant No.\,12-02-31430${\backslash}$12. 
Equipment of the Center of Collective Use of Scientific Equipment of A.~M.~Prokhorov General Physics Institute of RAS was utilized for this study.
We appreciate the support of this work.

We thank Ms.~Natalya V.~Kiryanova for her valuable contribution to arrangement and management of this research.

\end{acknowledgments}

\bibliography{Growing_Ge_hut}

\end{document}